\documentclass[journal,10pt, comsoc]{IEEEtran}
\usepackage[T1]{fontenc}

\usepackage{url}

\usepackage{cite}

\ifCLASSINFOpdf
  \usepackage[pdftex]{graphicx}
  \graphicspath{{Figure/}}
  \DeclareGraphicsExtensions{.pdf,.jpeg,.png}
\else
\fi
\usepackage{amsmath}
\usepackage[cmintegrals]{newtxmath}
\usepackage{algorithm}
\usepackage{algorithmic}
\usepackage{booktabs}
\usepackage{multirow}
\begin{document}
%
\title{Smart Name Lookup for NDN Forwarding Plane via Neural Networks}
%
%
%

\author{Zhuo~Li,~\IEEEmembership{Member,~IEEE,}
        Jindian~Liu,
        Liu~Yan,
        Beichuan~Zhang,~\IEEEmembership{Member,~IEEE,}
        Peng~Luo,
        Kaihua~Liu,~\IEEEmembership{Member,~IEEE}

\thanks{This work was supported by the National Natural Science Foundation of China under Grant 61602346, the Key R \& D projects of Hebei Province under Grant 20314301D, Tianjin Science and Technology Plan Project under Grant 20JCQNJC01490 and the Independent Innovation Fund of Tianjin University under Grant 2020XRG-0102. \textit{(Corresponding author: Zhuo Li, Jindian Liu, Liu Yan)}}
\thanks{Z. Li, J. Liu, L. Yan, K. Liu are with the school of Microelectronics, Tianjin University, and Tianjin microelectronics technology key laboratory of imaging and perception, Tianjin 300072, China (email: zli@tju.edu.cn; liujindian@tju.edu.cn; yanliu@tju.edu.cn; liukaihua@tju.edu.cn).}
\thanks{B. Zhang is with the Computer Science Department, University of Arizona, Tucson, AZ 85721, US (email: bzhang@cs.arizona.edu).}
\thanks{P. Luo is with the State Grid Hebei Electric Power Research Institute, ShijiaZhuang 050021, China (email: luopeng1984@sohu.com).}}

\maketitle


\begin{abstract}
  Name lookup is a key technology for the forwarding plane of content router in Named Data Networking (NDN). To realize the efficient name lookup, what counts is deploying a high-performance index in content routers. So far, the proposed indexes have shown good performance, most of which are optimized for or evaluated with URLs collected from the current Internet, as the large-scale NDN names are not available yet. Unfortunately, the performance of these indexes is always impacted in terms of lookup speed, memory consumption and false positive probability, as the distributions of URLs retrieved in memory may differ from those of real NDN names independently generated by content-centric applications online. Focusing on this gap, a smart mapping model named Pyramid-NN via neural networks is proposed to build an index called LNI for NDN forwarding plane. Through learning the distributions of the names retrieved in the static memory, LNI that will be trained by real NDN names offline and preset in content routers in the future can not only reduce the memory consumption and the probability of false positive, but also ensure the performance of real NDN name lookup. Experimental results show that LNI-based FIB can reduce the memory consumption to 58.258 MB. Moreover, as it can be deployed on SRAMs, the throughput is about 177 MSPS, which well meets the current network requirement for fast packet processing.
\end{abstract}




%
\IEEEpeerreviewmaketitle

\section{Introduction}
    \IEEEPARstart{N}{amed} Data Networking (NDN) \cite{zhang2010named} is proposed as an entirely new network architecture for future Internet, in which packets carry data names rather than IP addresses. In NDN, all communications are driven by the receiving end, i.e., the consumers \cite{zhang2010named}, through the exchange of two distinct types of packets: Interest and Data\cite{zhang2014named}. Both types of packets carry a name, which identifies a piece of data that can be transmitted in one Data packet. To fetch desired content, a consumer sends out an Interest packet with a unique identifying name to the network. Routers use this name to forward the Interest towards the producer(s) \cite{zhang2010named}. On the forwarding path, once the Interest reaches a node that has the requested Data, i.e., the Interest name is the same with the Data name or a prefix of the Data name \cite{cheng2012smart}, the Data containing both the name and the content will follow the reversed path taken by the Interest to get back to the consumer \cite{Li2018Packet}. 

    For the name-based packet forwarding, the core component is the stateful forwarding plane \cite{carofiglio2015pending}, where three tables including Content Store, Pending Interest Table (PIT) and Forwarding Information Base (FIB) are deployed \cite{zhang2010named}. Content Store is a temporary cache of Data packets the router has received to answer re-requested Interest packets. PIT stores all the Interests that a router has forwarded but not satisfied yet where each entry records a name carried in an Interest packet, together with its incoming and outgoing interfaces. FIB stores a set of name prefixes of Interest packets announced in routing and its outgoing interfaces as the next hop to ensure proper packet forwarding. 

  \begin{table}[!t]
    \renewcommand{\arraystretch}{1.2}
    \newcommand{\tabincell}[2]{\begin{tabular}{@{}#1@{}}#2\end{tabular}}
    \centering
    \caption{Number of chains with different length: Blacklist-2013 vs. Blacklist-2020}
    \label{Table1_Number of collision: Blacklist-2013 vs. Blacklist-2020}
    \setlength{\tabcolsep}{3mm}{
    \begin{tabular}{ccccccc}
    \hline
    \multirow{2}{*}[-0.8ex]{Dataset} & \multirow{2}{*}[-0.8ex]{\# of URLs} & \multicolumn{5}{c}{\# of chains with different length}              \\ \cmidrule(l){3-7} 
                             &                             &  2 &  3 &  4 &  5 &  6 \\ \hline
    Blacklist-2013           & 20,000                       & 3,733        & 1,217        & 283         & 70          & 8           \\
    Blacklist-2020           & 20,000                       & 3,658        & 1,207        & 327         & 60          & 6           \\ \hline
    \end{tabular}}
    \end{table}

  \begin{figure}[!t]
        \centering
        \includegraphics[width=2.55in]{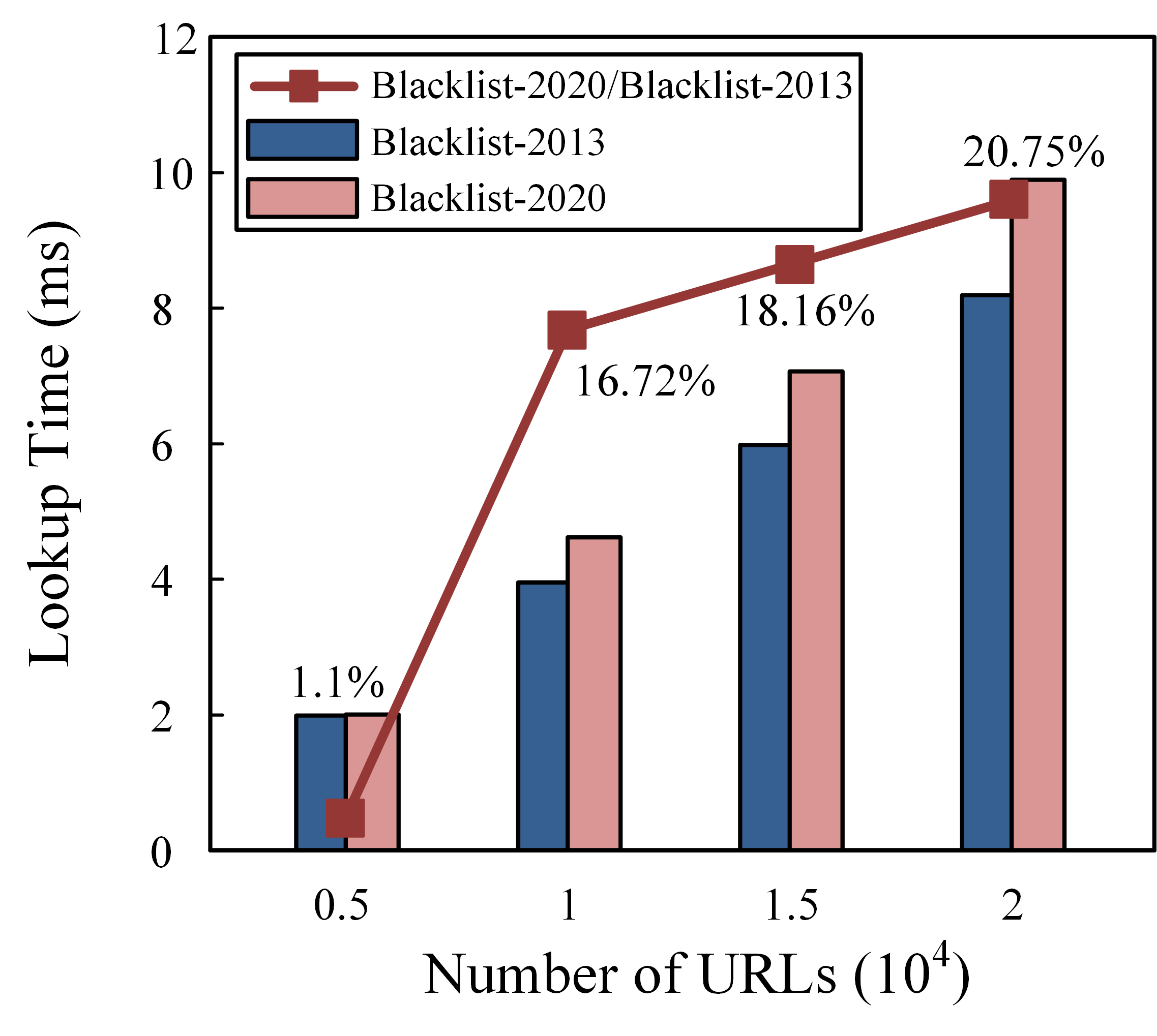}
        \caption{Lookup time (ms): Blacklist-2013 vs. Blacklist-2020.}
        \label{Figure7_Distribution of mapped slots in memory}
    \end{figure}

  Given the forwarding plane in NDN is so different from IP, it has to imply a substantial re-engineering lookup structures for fast, memory-efficient, and scalable packet forwarding. So far, many novel indexes based on trie \cite{fredkin1960trie}, hash table \cite{zink2009survey}, Bloom filter \cite{bloom1970space} and skip list \cite{Pugh:1990:SLP:78973.78977} are proposed for NDN forwarding plane to support efficient name lookup. As the large-scale real NDN names are not available yet, most of these indexes are optimized for or evaluated with URLs collected from the current Internet. Unfortunately, the performance of indexes is always impacted by the distributions of data retrieved in the memory, in terms of lookup speed, memory consumption and false positive probability. As the distributions of URLs retrieved in the index of NDN forwarding plane may differ from those of real NDN names independently generated by content-centric applications online, the performance of these indexes designed with URLs will be degraded. To clearly illustrate this impact, we extracted two URL datasets from Backlist \cite{blacklist} in 2013 and 2020, and tested their position located in the memory by CityHash. The pre-experimental result in Table I shows that the distributions of URLs located in the static memory between the two datasets are different. That is the chains to deal with conflicts differ in number and length, which certainly affects both of the lookup speed and memory consumption. As shown in Fig. 1, the gap between the lookup time of the two datasets increases due to the different number of chains utilized. Given that more than 1 million names have to be stored in NDN forwarding plane, the performance of name lookup will be more seriously impacted by the distribution of real names in the memory. To ensure the performance of NDN forwarding plane, the proposed indexes have to be redesigned, which leads to the high engineering effort. Therefore, it is crucial to design an index that can adapt to the distributions of names retrieved in the memory.

  To tackle this gap, a smart mapping model based on neural networks, called Pyramid-NN, is proposed to construct a mapping function that can adapt to the distributions of real NDN names retrieved by learning the distributions in the static memory. Moreover, based on Pyramid-NN, an index called LNI is proposed. Through learning the distributions of names in the static memory, LNI that will be trained by real NDN names offline and preset in content routers in the future can not only reduce the memory consumption and the probability of false positive, but also ensure the performance of real name lookup. The main contributions are as follows:

    \begin{enumerate}
      \item A smart pyramid-like neural network model named Pyramid-NN is first conceived to build an efficient index that can adapt to data distributions. Pyramid-NN learns the distributions of data retrieved in the static memory by training, so that it can not only adapt to the distributions of data retrieved well, but also map the data more uniformly, which improves the memory utilization. The architecture of Pyramid-NN is designed to be a multi-level model consisting of a number of Back Propagation Neural Networks (BPNNs) \cite{rumelhart1986learning}, and the level-by-level training algorithm of it is put forward to support efficient model training. Moreover, the proper hyperparameters of it are selected by simulations and analysis.

      \item Based on Pyramid-NN, an index called Learning Name Index (LNI) for NDN forwarding plane is proposed to support the efficient name lookup. In LNI, the Input Processor turns variable-length NDN name to fixed-dimensional input vector; Pyramid-NN does the mapping, which can improve the memory utilization, as well as adapt to complex NDN names as it can collect training data and select label rules flexibly; the Enhanced Bitmap \cite{Li20185G} gets the memory address for storing data, which further reduces the memory consumption. Moreover, the lookup algorithm of LNI is also proposed to implement fast name lookup.

      \item The performance of Pyramid-NN is presented in detail to verify the feasibility for the mapping. As Pyramid-NN can learn the distributions of data retrieved in the static memory, compared with the traditional hash functions with the probability of false positive under 1\%, Pyramid-NN requires only about 25\% of the slots and the execution speed is on the same order of magnitude as that of the CityHash256.

      \item The performance of LNI-based FIB, called LNI-FIB, is evaluated and discussed by executing contrast experiments with three state-of-the-art indexes, namely Hash Table-FIB, Binary Patricia Trie-FIB and B-MaFIB. The results show that LNI-FIB extremely reduces the memory consumption to 58.258 MB with the probability of false positive under 1\%, which means it can easily fit into contemporary SRAMs in commercial line card. Also, the throughput of LNI-FIB is about 177 million searches per second (MSPS), which well meets the current network requirement for fast packet processing.
    \end{enumerate}

	The remainder of this paper is organized as follows. Section II surveys related work. Section III provides the design essentials of NDN name lookup. Section IV presents Pyramid-NN, including the design overview, the model architecture, the training process and the model hyperparameter selection. Section V describes LNI, presenting its architecture and the process of lookup. Section VI shows the performance of Pyramid-NN in detail. Section VII compares and discusses the performance of LNI-FIB by executing contrast experiments. Section VIII gives a brief conclusion and future work.

\section{Related Work}
  In this section, the indexes proposed in NDN forwarding plane are summarized \cite{Li2018Packet}, which are classified into five types, namely trie-based, hash table-based, Bloom filter-based, skip list-based and machine learning-based.
  
\subsection{Trie-based Schemes}
    The logical characteristics of trie can reduce the memory consumption of the hierarchical names stored in NDN router, so \cite{Ghasemi2018A, dai2016consert, lee2016new, song2015scalable, liu2017unified, Yuan2015Data, Seo2018Bitmap, afanasyev2016nfd, saxena2016reliable, saxena2016radient, li2016improved, saxena2016n, bouk2015hierarchical, Feng2015A, quan2014tb2f, wang2013wire, wang2012scalable, dai2012pending, wang2011parallel} propose trie-based schemes for NDN forwarding plane. Among them, the main research issues are how to design its granularity to reduce the memory consumption and how to reduce its depth to improve the lookup speed. For example, NameTrie \cite{Ghasemi2018A} proposes minASCII encoding to store and index forwarding information more efficiently; CONSERT \cite{dai2016consert} removes the redundancy to minimize the number of name prefixes; PC-NPT \cite{lee2016new} proposes path compression to reduce the average number of node accesses; Binary Patricia Trie \cite{song2015scalable} uses binary as the granularity to minimize the impact of redundant information at memory; CTrie \cite{liu2017unified} builds a combinational trie structure from both component-based and byte-based hierarchical names to achieve the unified index. 

\subsection{Hash Table-based Schemes}
    Hash table has advantages in lookup speed, so \cite{Yuan2015Data, yuan2015reliably, shen2018code, hu2019fast, varvello2013design, yuan2012scalable, thomas2015object, yuan2014scalable, so2013named, wang2013greedy, so2012toward} propose hash table-based schemes for NDN forwarding plane. Among them, the main research issues are how to ensure more accurate forwarding, how to reduce the memory consumption and how to support the algorithms of name matching. For example, FHT \cite{Yuan2015Data} uses fingerprint collision table to reduce the memory consumption; Binary Search of Hash Tables \cite{yuan2015reliably} constructs a balanced binary search tree to improve the execution efficiency of name matching algorithm; CoDE \cite{shen2018code} achieves fast name lookup and update using conflict-driven encoding; MOBS \cite{hu2019fast} concentrates on the optimization of random search algorithm to reduce the memory consumption. However, the proposed schemes have to store all the content names additionally to ensure accurate forwarding, which causes memory inefficient.

\subsection{Bloom Filter-based Schemes}
    Bloom filter can greatly reduce the memory consumption, so \cite{wang2013namefilter, you2012dipit, dai2016bfast, li2017hybrid, li2014mapit, Li20185G, Hou2018Bloom, Munoz2017I, Zhang2017Adaptive, lee2016name, shimazaki2016hash, manghnanilength, yu2016hardware, perino2014caesar, quan2014scalable, fukushima2013efficient, li2012compression} propose Bloom filter-based schemes for NDN forwarding plane. Among them, the main research issue is how to solve the problem that Bloom filter can only determine whether an element is in the set but not locate its memory address. For example, NameFilter \cite{wang2013namefilter} and DiPIT \cite{you2012dipit} assign a Bloom filter to each interface of the forwarding plane; BFAST \cite{dai2016bfast} combines Bloom filter with hash table; MBF \cite{li2017hybrid, li2014mapit} and B-MBF \cite{Li20185G} combines Bloom filter with Mapping Array and Bitmap to locate the memory address. 

\subsection{Skip List-based Schemes}
    Skip list can preserve the order of data storage and effectively support the cache replacement policy, so \cite{yuan2012scalable} and \cite{pan2016fast} propose skip list-based schemes for NDN forwarding plane. For example, Locality-Aware Skip List \cite{pan2016fast} records the address of skip list node accessed when querying a node, which improves the lookup speed to some extent. However, the time complexity of such schemes is generally high due to the limitation of its basic structure \cite{Li2018Packet}.

\subsection{Machine learning-based Schemes}
    In view of the rapid development of machine learning techniques in recent years, machine learning-based schemes have been proposed in indexes for NDN forwarding plane. Learning Tree \cite{yan2019learning} was posted to learn the distribution of data to build an efficient index. Learned Bloom-Filter Lookup\cite{wang2019learned} combines Recursive Neural Network (RNN) with standard Bloom filter to improve lookup efficiency.
    
\subsection{Summary}
    The indexes based on traditional data structure mentioned above have shown good performance, most of which are optimized for or evaluated with URLs collected from the current Internet, as the large-scale real NDN names are not available yet. However, the distributions of real NDN names may differ from that of URLs used currently, which will degrade the performance of the indexes, as the results shown in the pre-experiments in Section I. Therefore, the indexes have to be redesigned to ensure the performance. Fortunately, machine learning brings the opportunity to tackle this issue. Therefore, an index via neural networks, called LNI, is proposed in this paper. Through learning the distributions of the names retrieved in the static memory, LNI that will be trained by real NDN names offline and preset in content routers in the future can not only reduce the memory consumption and the probability of false positive, but also ensure the performance of real name lookup.

    \section{Design Essentials of NDN Name Lookup}

    \begin{figure*}[!t]
        \centering
        \includegraphics[width=6.5in]{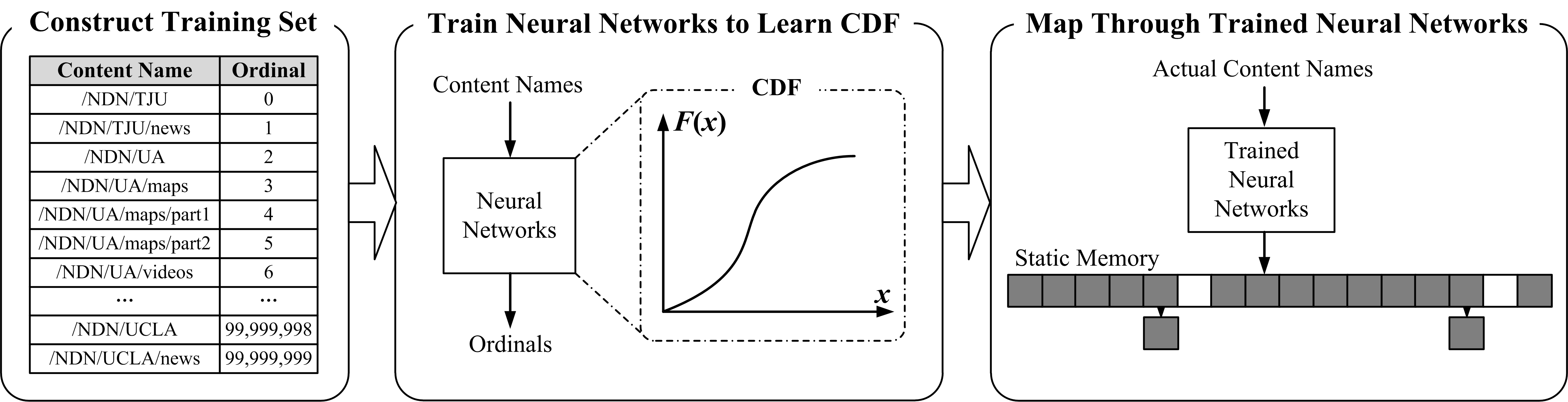}
        \caption{Design overview of Pyramid-NN.}
        \label{Design overview of Pyramid-NN}
    \end{figure*}
       
   In this section, by summarizing the existing theoretical results about the NDN name lookup, three design essentials of NDN name lookup are described in detail respectively.

    \subsection{Complex Name Structure} Unlike IP addresses of fixed length, NDN names are variable-length with no upper bound, having complex and unrestrained formats. And what's worse, so far the design tussle of data naming remains an open challenge due to the different requirements that applications, security, and the network place on data names \cite{jacobson2017named}.  

	As lookup keys in NDN forwarding plane, the complex NDN names have to be scanned in the forwarding processes. Consequently, NDN forwarding plane has to support effective lookup for arbitrary complex names.
	
    \subsection{Small Memory Footprint} Compared with IP, NDN forwarding plane calls for much more memory space for two reasons. First, the number of entries in NDN forwarding plane is orders-of-magnitude greater than that in IP. Taking PIT as an example, since an Interest stays in PIT of each NDN router along the path until the corresponding Data returns, PIT needs to hold 1 million entries for 10 Gbps gateway trace and 1.5 million entries for 20 Gbps at least \cite{yi2012adaptive}. Second, the size of each entry in NDN forwarding plane is also larger, for an NDN name is more complex than an IP address. These factors together result in the forwarding tables with larger memory footprint than IP-forwarding tables. Therefore, it is really a great challenge to study how to reduce the mempry consumption of the indexes for three tables in NDN forwarding plane, so that they can be deployed in small and high-speed memories (e.g., SRAM).

    \subsection{High Throughput} NDN forwarding plane requires quite frequent name lookup and update operations \cite{Li2018Packet}. Whenever Interest/Data(s) arrive at NDN routers or the routing protocols recompute FIB, the corresponding names have to be scanned and corresponding operations have to be performed in NDN forwarding plane. As reported in \cite{varvello2013design}, considering a load equal to 100\%, PIT’s operations peak at 60 million per second; in a more realistic scenario with flow balance or a load of 50\%, the frequency of PIT’s operations is about 6 million per second. Therefore, NDN forwarding plane has to perform name lookup at a practicable high speed, so that it can satisfy the requirement for high throughput in NDN router.
    
    \section{Smart Mapping Model: Pyramid-NN}
    Learning techniques are introduced to design a smart mapping model via neural networks, called Pyramid-NN. In this section, the design overview, the model architecture, the training process and the model hyperparameter selection are described in detail.
  
  \subsection{Design Overview}
     In order to build an index that can adapt to data distributions in the memory, Pyramid-NN uses neural networks to learn the distributions of the data retrieved in the static memory. The distributions of the data retrieved are reflected in its cumulative distribution function (CDF), whose value represents the likelihood of a key less than or equals to the lookup key. The property of the CDF states that for a data set with arbitrary distribution the values calculated by its CDF have a uniform distribution on [0, 1]. Utilizing the CDF as a mapping function, the probability of each key mapped to different slots is the same. Therefore, the slots mapped can be uniformly distributed through multiplying the values of the CDF by the total number of slots in memory.
  
     Specifically, the design overview of Pyramid-NN is illustrated in Fig. 2. The first phase is to construct the training set. A large number of variable-length content names are collected and turned into fixed-dimensional vectors. These vectors, as lookup keys in NDN route table, are sorted based on the values of them, then labeled with the ordinals. The second phase is to train neural networks using the vector-ordinal pairs to learn the CDF of the data retrieved. The final phase comes into application, which is to do the mapping through trained neural networks. In practice, the Pyramid-NN will be trained with real NDN names offline and deployed in content routers. The names of NDN packets are input into the trained Pyramid-NN and then the CDF values are estimated. Finally, the mapped slots are obtained by multiplying the values of the CDF by the total number of slots in memory, which can be distributed more uniformly.
  
   \begin{figure}[!t]
          \centering
          \includegraphics[width=3.0in]{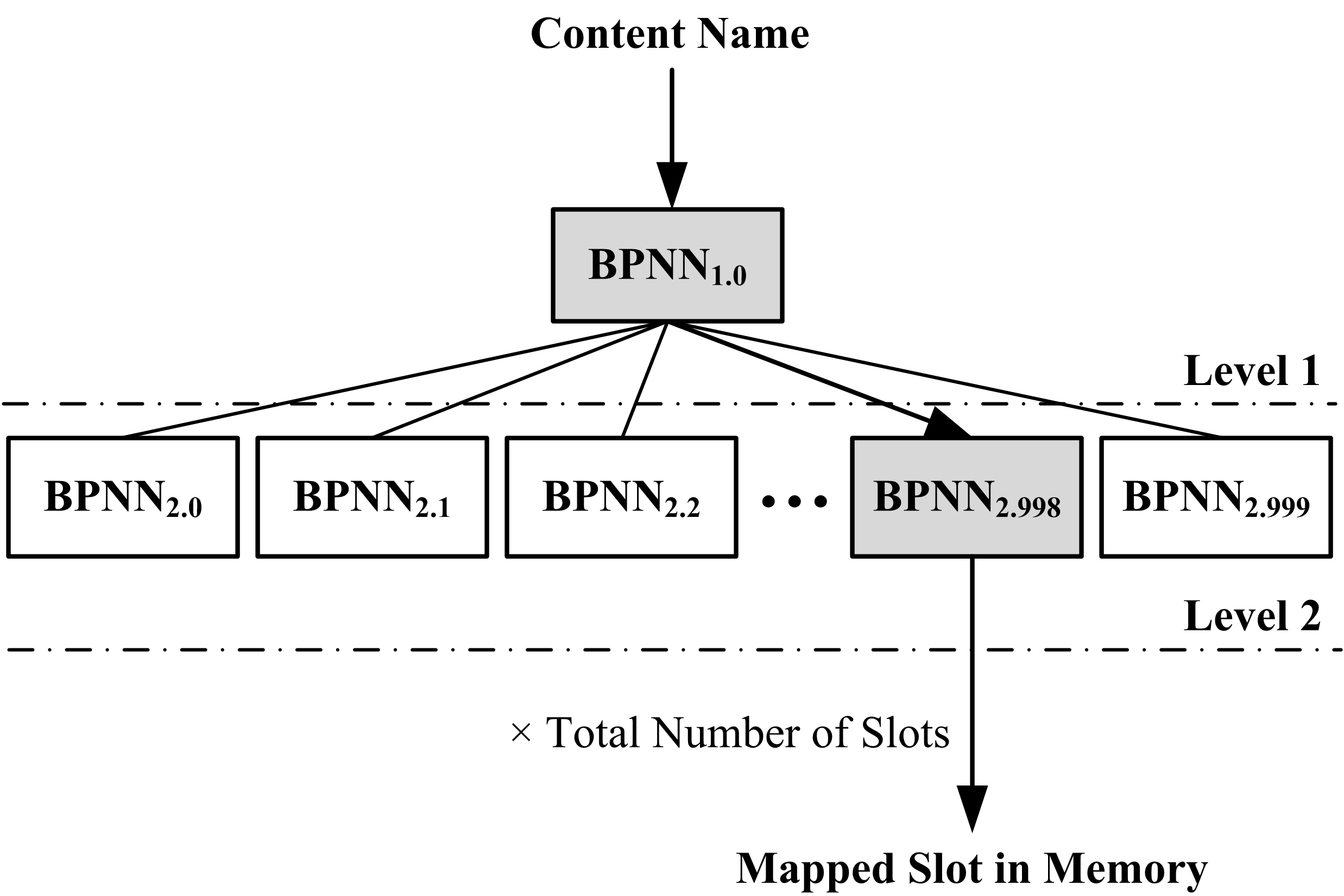}
          \caption{Model architecture of Pyramid-NN.}
          \label{Model architecture of Pyramid-NN}
      \end{figure}
  
  \subsection{Model Architecture and Training Process}
      Lookup speed is the most critical requirement for the index. Therefore, BPNNs are used to build Pyramid-NN, as it has strong parallel computing ability \cite{xu2017continuous}. However, the classification accuracy of a single BPNN (i.e., the number of model level is 1) is only 0.18\%, as shown in Fig. 4(a). Therefore, it is difficult to accurately learn the CDF of millions of data. But when there are 2 or 3 levels, Pyramid-NN can get pretty good accuracy performance, as it can efficiently divide the large namespace into multiple smaller sub-namespaces so that each BPNN at the last level can accurately represent the CDF of relatively little data. Thus, Pyramid-NN is designed to be a multi-level neural network model. Meanwhile, the multi-level model can work in parallel to further improve the lookup speed.
  
      Fig. 3 gives an example of two-level Pyramid-NN, which consists of 1 BPNN at level 1 and 1,000 BPNNs at level 2. Suppose BPNN$_{\textit{j}.\textit{k}}$ represents the \textit{k}-th BPNN at level \textit{j}. BPNN$_{1.0}$ is trained to get the region number from 0 to 999, each of which corresponds to a BPNN$_{2.\textit{k}}$ (0 $\le$ \textit{k} $\le$ 999). Each BPNN$_{2.\textit{k}}$ is trained to learn a part of the CDF. Therefore, the estimation range of all the trained BPNN$_{2.\textit{k}}$ can cover the entire CDF, i.e., the trained Pyramid-NN can be seen as a function that estimates the CDF value.
  
    \begin{algorithm}[!t]
      \caption{Training process of Pyramid-NN}
      \begin{algorithmic}[1]
      \STATE \textbf{Procedure} Training (\textit{Training\_Names}[])
      \STATE \textit{Size} $\leftarrow$ \textit{Training\_Names}.\textit{size};
      \FOR{each \textit{i} $\in$ [0, \textit{Size} - 1]}
          \STATE Inputs[\textit{i}][:] $\leftarrow$ \textit{Input\_Process}(\textit{Training\_Names}[\textit{i}]);\
          \STATE L1\_Labels[\textit{i}] $\leftarrow$ $\lfloor$\textit{i} / \textit{Size} $\times$ 1000$\rfloor$;\
          \STATE L2\_Labels[\textit{i}] $\leftarrow$ \textit{i} / \textit{Size};\
      \ENDFOR
      \STATE \textit{Sort}(Inputs);
      \FOR{each \textit{j} $\in$ [1, \textit{max\_epochs}]}
          \STATE \textit{L1\_Outputs} $\leftarrow$ \textit{\rm{BPNN}$_{1.0}$}(Inputs);\
          \STATE \textit{Loss} $\leftarrow$ \textit{Loss Function}(L1\_Outputs, L1\_Labels);\
          \STATE \textit{Backward}(\textit{Loss});\
          \STATE \textit{Update}(Parameters);\
      \ENDFOR
      \STATE \textit{Save}(\textit{\rm{BPNN}$_{1.0}$});
      \FOR{each \textit{k} $\in$ [0, 999]}
          \STATE \textit{index} $\leftarrow$ \textit{Find\_Rows}(\textit{L1\_Outputs} = \textit{k});\
          \FOR{each \textit{j} $\in$ [1, \textit{max\_epochs}]}
               \STATE\textit {L2\_k\_Outputs} $\leftarrow$ \textit{\rm{BPNN}$_{2.k}$}(Inputs[\textit{index}][:]);\
             \STATE \textit{Loss} $\leftarrow$ \textit{Loss Function}(\textit{L2\_k\_Outputs}, \\ \qquad \qquad \qquad \qquad \qquad L2\_Labels[\textit{index}]);\
               \STATE \textit{\rm{BPNN}$_{2.k}$} $\leftarrow$ \textit{Train}(Inputs[\textit{index}][:], \\ \qquad \qquad \qquad \qquad L2\_Labels[\textit{index}]);\
             \STATE \textit{Backward}(\textit{Loss});\
               \STATE \textit{Update}(Parameters);\
          \ENDFOR
          \STATE \textit{Save}(\textit{\rm{BPNN}$_{2.k}$});
      \ENDFOR
      \STATE \textbf{End Procedure}
      \end{algorithmic}
      \end{algorithm}
  
      Algorithm 1 shows the training process of Pyramid-NN. First, for each \textit{i} $\in$ [0, \textit{Size} - 1] where \textit{Size} represents the number of training names, the \textit{i}-th name is processed to the corresponding input vector (line 4). The detail process is as follows. Suppose the input vector is \textit{y} with dimension \textit{N}, and an \textit{n}-length NDN name can be seen as a vector \textit{x} $\in$ $\mathbb{R}$$^{\textit{n}}$, where \textit{x}$_{i}$ is the ASCII value of the \textit{i}-th character. For a name with length \textit{n} $\le$ \textit{N}, set \textit{y}$_{\textit{i}}$ = \textit{x}$_{\textit{i}}$ where 0 $\le$ \textit{i} $<$ \textit{n} and \textit{y}$_{\textit{i}}$ = 0 where \textit{n} $\le$ \textit{i} $<$ \textit{N}. For \textit{n} $>$ \textit{N}, split every \textit{N} elements of \textit{x} into a sub-vector, then set \textit{y}$_{\textit{i}}$ = \textit{Bitxor(the i-th elements in all sub-vectors)} where 0 $\le$ \textit{i} $<$ \textit{N}, considering that bitxor method can achieve faster processing speed than hash function. Thus the \textit{N}-dimensional input vector \textit{y} is got; $\lfloor$\textit{i} / \textit{Size} $\times$ 1000$\rfloor$ is calculated as the \textit{i}-th label of level 1, which is obviously an integer in [0, 999]; \textit{i} / \textit{Size} is calculated as the \textit{i}-th label of level 2, which is a decimal in [0, 1]. Then all the input vectors are sorted in ascending order to correspond to the labels (line 5-8). Second, \textit{\rm{BPNN}$_{1.0}$} is trained with all the input vectors and labels of level 1 (line 9-15). Third, all of the \textit{\rm{BPNN}$_{2.k}$}s are trained based on the outputs of level 1. For each \textit{k} $\in$ [0, 999], all the input vectors with output \textit{k} at level 1 and their corresponding labels of level 2 are picked out as the input vectors and labels of \textit{\rm{BPNN}$_{2.k}$} respectively, and then \textit{\rm{BPNN}$_{2.k}$} is trained (line 16-26).
  
      For each BPNN$_{\textit{j}.\textit{k}}$ in Pyramid-NN, the time complexity of training is O(n) where n is the epoches of training. The time complexity does not affect its performance, as Pyramid-NN is trained offline before being in content routers.
  
      \begin{figure}[!t]
          \centering
          \includegraphics[width=3.0in]{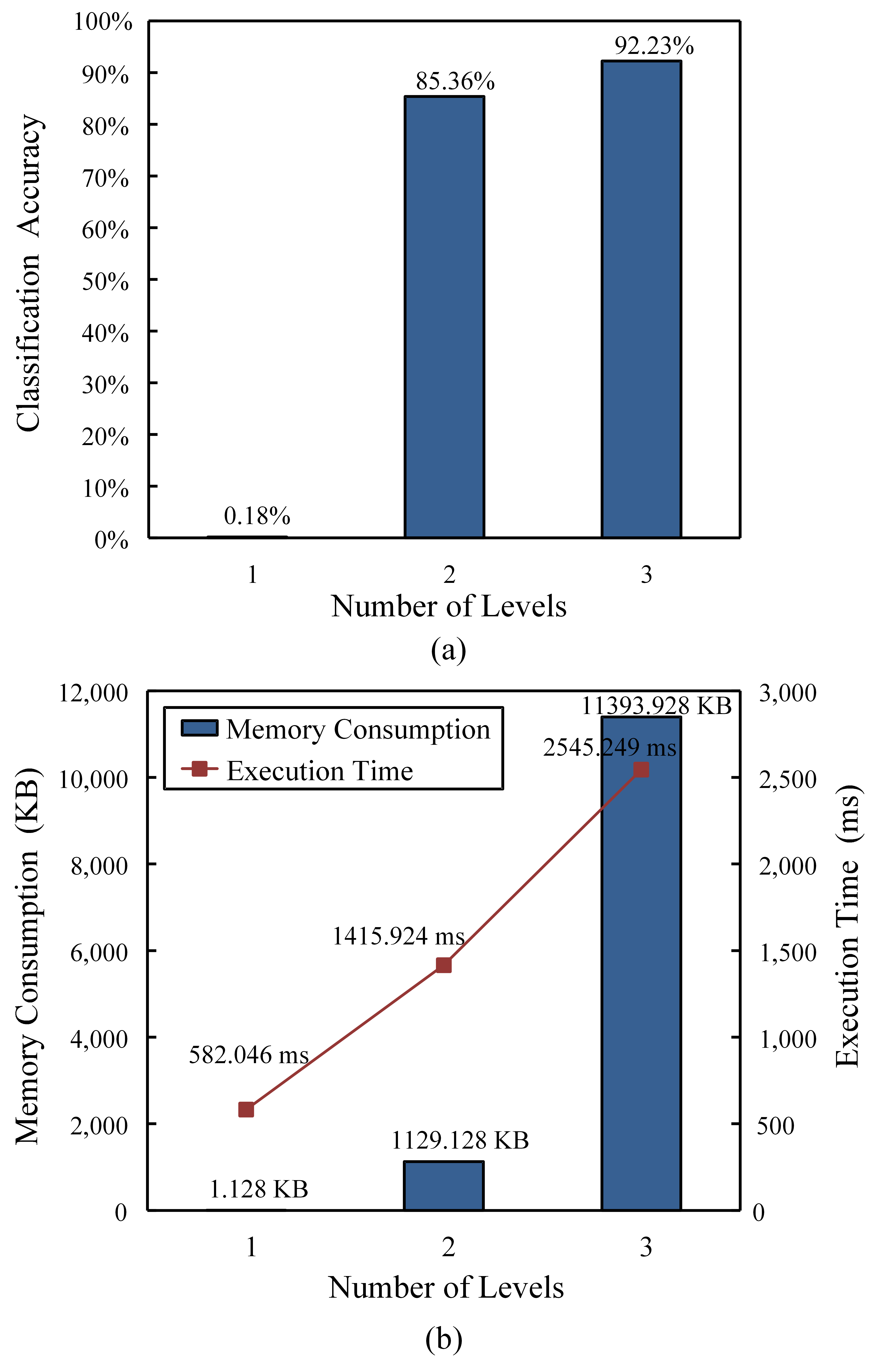}
          \caption{Performance of Pyramid-NN under different number of levels. (a) Classification accuracy performance. (b) Memory consumption and execution time performance.}
          \label{Performance of Pyramid-NN under different number of levels}
      \end{figure}
  
      \begin{table}[!t]
\renewcommand{\arraystretch}{1.3}
\newcommand{\tabincell}[2]{\begin{tabular}{@{}#1@{}}#2\end{tabular}}
\centering
\caption{Dataset of Domain Names}
\label{Table1_Dataset of Domain Names}
\setlength{\tabcolsep}{2.5mm}{
\begin{tabular}{cccc}
\hline
Dataset           & \# of Names & Average Length (B) & Size (MB) \\ \hline
Training Set      & 100,000,000 & 25.871             & 2,624.170 \\
Validation Set    & 2,000,000   & 25.880             & 52.511    \\
Testing Set 1     & 500,000     & 25.895             & 13.133    \\
Testing Set 2     & 1,000,000   & 25.864             & 26.235    \\
Testing Set 3     & 1,500,000   & 25.872             & 39.364    \\
Testing Set 4     & 2,000,000   & 25.887             & 52.514    \\ \hline
\end{tabular}}
\end{table} 
  
  \subsection{Model Hyperparameter Selection}
      It is generally known that the hyperparameter selection is important when designing a neural network \cite{wei2019neural}. In Pyramid-NN, the number of model levels and the number of neurons for one BPNN directly affect not only the classification accuracy(i.e., the proportion of the names whose corresponding outputs and labels are matched in all input names), but also the memory consumption and the execution time. Furthermore, the number of input neurons for one BPNN also affects the probability of input collision(i.e., the proportion of distinct names processed to the same input vector in all names). Therefore, to select the proper hyperparameters for Pyramid-NN, Pyramid-NN under different hyperparameters is trained with Deep Learning Toolbox in MATLAB. The performance is tested on a workstation with an Intel Xeon E5-1650 v2 CPU of 3.50 GHz and DDR3 SDRAM of 24 GB. Given that more than 1 million entries have to be stored in NDN forwarding plane \cite{Li2018Packet}, billions of different NDN names, based on the naming conventions\cite{yu2014ndn}, are generated as the dataset of experiments, where 100 million are used as the training set, 2 million as the validation set and 0.5-, 1-, 1.5-, 2-million as the testing sets, as listed in Table II.
  
      Note that the hyperparameters selected below only represent one possible solution for implementing Pyramid-NN, but does not mean that Pyramid-NN must use the below hyperparameters.

      \begin{figure}[!t]
          \centering
          \includegraphics[width=3.0in]{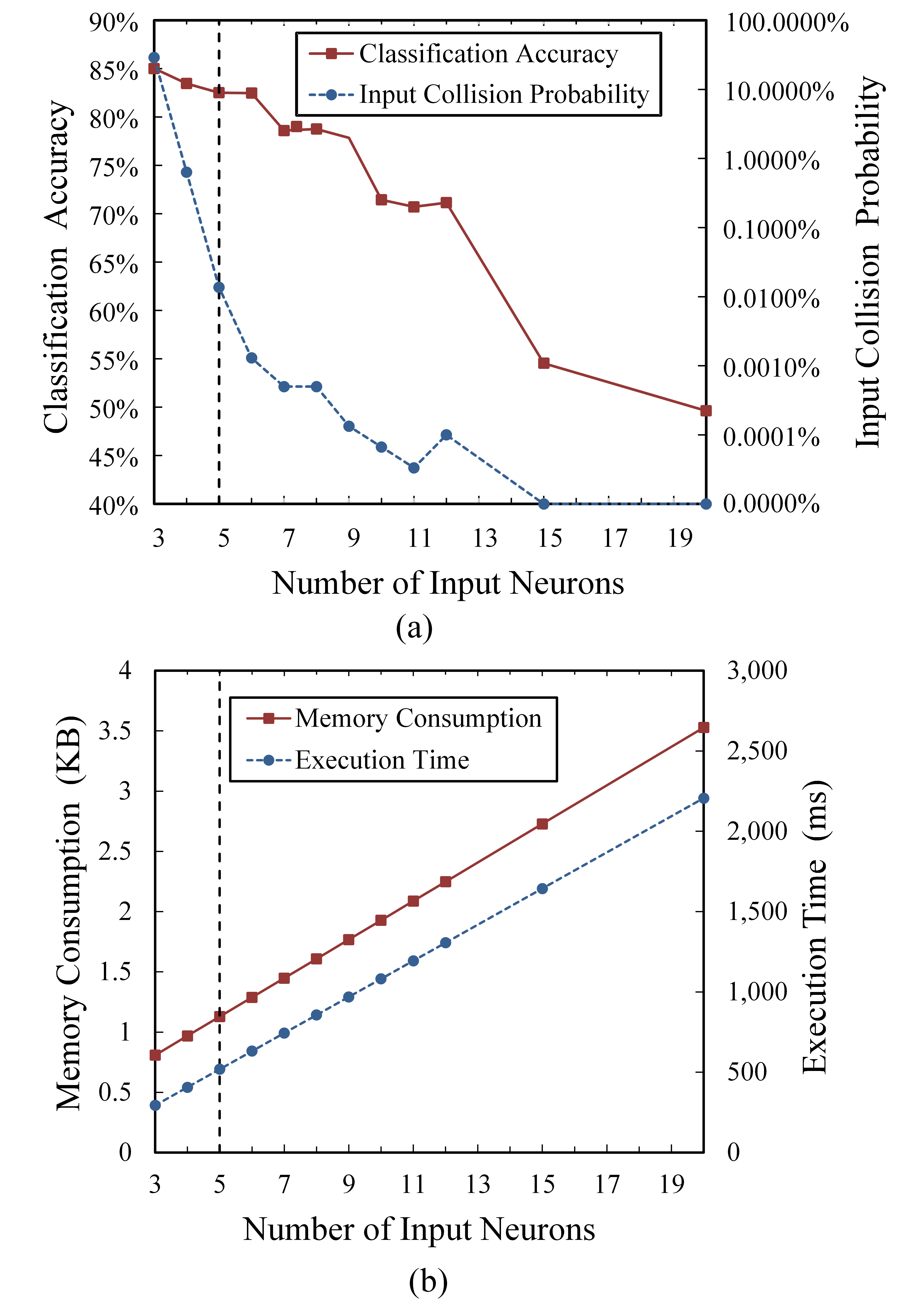}
          \caption{Performance of the BPNN under different number of input neurons. (a) Classification accuracy and input collision probability performance. (b) Memory consumption and execution time performance.}
          \label{Performance of the BPNN under different number of input neurons}
      \end{figure}
  
      \begin{figure}[!t]
          \centering
          \includegraphics[width=3.0in]{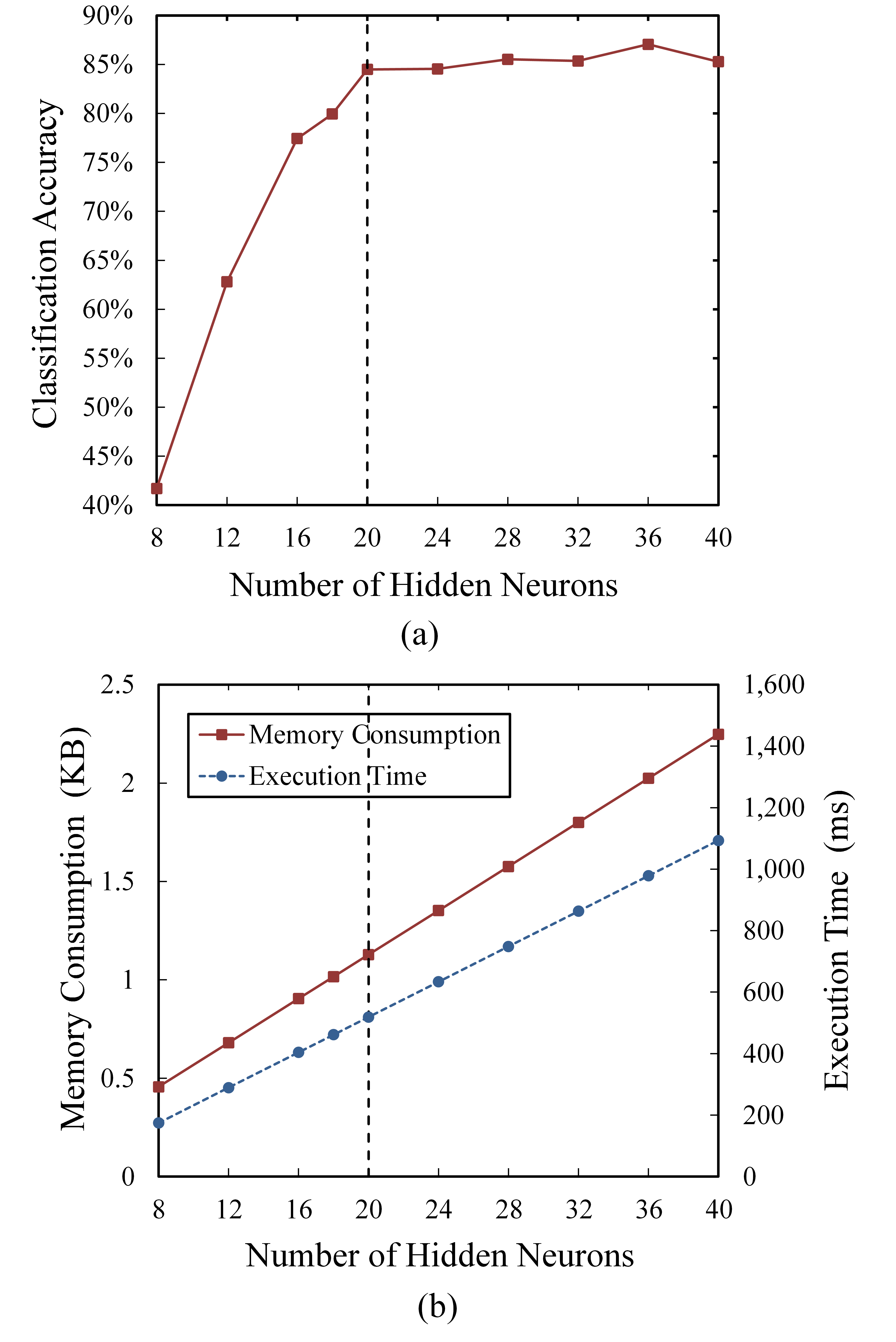}
          \caption{Performance of the BPNN under different number of hidden neurons. (a) Classification accuracy performance. (b) Memory consumption and execution time performance.}
          \label{Performance of the BPNN under different number of hidden neurons}
      \end{figure}
  
  \subsubsection{Number of Model Levels}
      The classification accuracy of Pyramid-NN is acceptable when there are 2 or 3 levels, as shown in Fig. 4(a). Moreover, fig. 4(b) shows the memory consumption of Pyramid-NN and the time consumption per million executions, where the 2-level Pyramid-NN requires less memory and executes relatively fast compared with the 3-level Pyramid-NN, so we determine Pyramid-NN to be a two-level model.
  
  \subsubsection{Number of  BPNNs at Level 2}
    As the number of entries in NDN forwarding plane is in the order of millions, 1,000 BPNNs are implemented at level 2, where each BPNN maps thousands of data to achieve more uniform mapping.
  
  \subsubsection{Number of Input Neurons for one BPNN}
       As illustrated in Fig. 5(a), with the increase of the number of input neurons, both of input collision probability and classification accuracy decrease. According to the algorithm of input process introduced in subsection IV(B), with the increase of the number of input neurons, the probability of distinct names processed to the same input vectors (i.e., the input collision probability) reduces, but the model will be overfitted, which leads to the lower classification accuracy.
  
       Specifically, when the number of input neurons is 3 to 6, the classification accuracy is greater than 80\%. However, as it is 5 and 6, the input collision probability can be much less than the other two cases, which meets the current network requirements for 1\% of packet loss rate well \cite{you2012dipit}. Therefore, the number of input neurons can be 5 or 6. Moreover, considering the memory consumption and execution time shown in Fig. 5(b), Pyramid-NN with 5 input neurons requires less memory and shorter execution time relatively compared with the 6-input neurons Pyramid-NN. Thus, the  number of input neurons is determined to be 5.
  
  \subsubsection{Number of Hidden Neurons for one BPNN}
      Fig. 6(a) shows that when the number of hidden neurons is greater than or equal to 20, the classification accuracy is stabilized at about 85\%. Further considering memory consumption and execution time shown in Fig. 6(b), Pyramid-NN requires more the memory consumption and longer execution time, as the number of hidden neurons increases. Hence the number of hidden neurons is determined to be 20.
  
      \section{Learning Name Index: LNI}
      Based on Pyramid-NN, an index is proposed for NDN forwarding plane, called LNI, which not only can support efficient NDN name lookup, but also adapt to the distributions of real NDN names in the static memory. In this section, the index architecture and lookup process of LNI are presented in detail.
 
 \subsection{Index Architecture}
 
     The index architecture of LNI is shown in Fig. 7, which contains three units: the Input Processor, Pyramid-NN and the Enhanced Bitmap.
 
   The Input Processor turns variable-length NDN name to fixed-dimensional input vector to support efficient lookup for complex names, as described in Subsection IV(B).
 
     Pyramid-NN does the mapping. According to the hyperparameter selection in Subsection IV(C), Pyramid-NN is a two-level model consisting of 1 BPNN at level 1 and 1,000 BPNNs at level 2, where each BPNN has 5 input neurons, 20 hidden neurons and 1 output neuron. Through the trained BPNNs, the mapped slot is got by multiplying the value of the CDF by the total number of slots.
 
     The Enhanced Bitmap gets the memory address for storing data, which further reduces the memory consumption \cite{Li20185G}. Specifically, the Enhanced Bitmap is composed of slots with 2 bytes each, and evenly segmented into 1000 equal parts. Meanwhile, each part corresponds to a BPNN in level 2 of Pyramid-NN and a memory space respectively. From the output of Pyramid-NN, the corresponding part of the Enhanced Bitmap is got and the mapped slot in this part is calculated. Then, the calculation result is inserted in this slot as the offset address. Based on the base address of memory space corresponding to this part and the offset address represented by the number in this slot, the actual memory address for storing forwarding information can be obtained.
 
     Obviously, the classification accuracy of the BPNN in Pyramid-NN and the number of slots in the Enhanced Bitmap jointly affect the false positive probability of LNI. For instance, if the name that is input into \textit{\rm{BPNN}$_{1.0}$} is mapped to a fault BPNN at level 2, this name will be mapped to the same slot in the Enhanced Bitmap occupied by the name classified correctly(i.e., the false positive), that is, the probability of false positive increases as the classification accuracy reduces. For the number of slots in the Enhanced Bitmap, the larger the number of slots, the more dispersed the data is in the Enhanced Bitmap, thus the false positive probability is reduced. With the classification accuracy of the BPNN in Pyramid-NN stabilized at about 85\% (as shown in Fig. 6(a)), the number of slots in the Enhanced Bitmap will be appropriately enlarged to reduce the false positive probability to less than 1\% which is the current network requirements for packet loss rate.
 
  \begin{figure}[!t]
         \centering
         \includegraphics[width=3.2in]{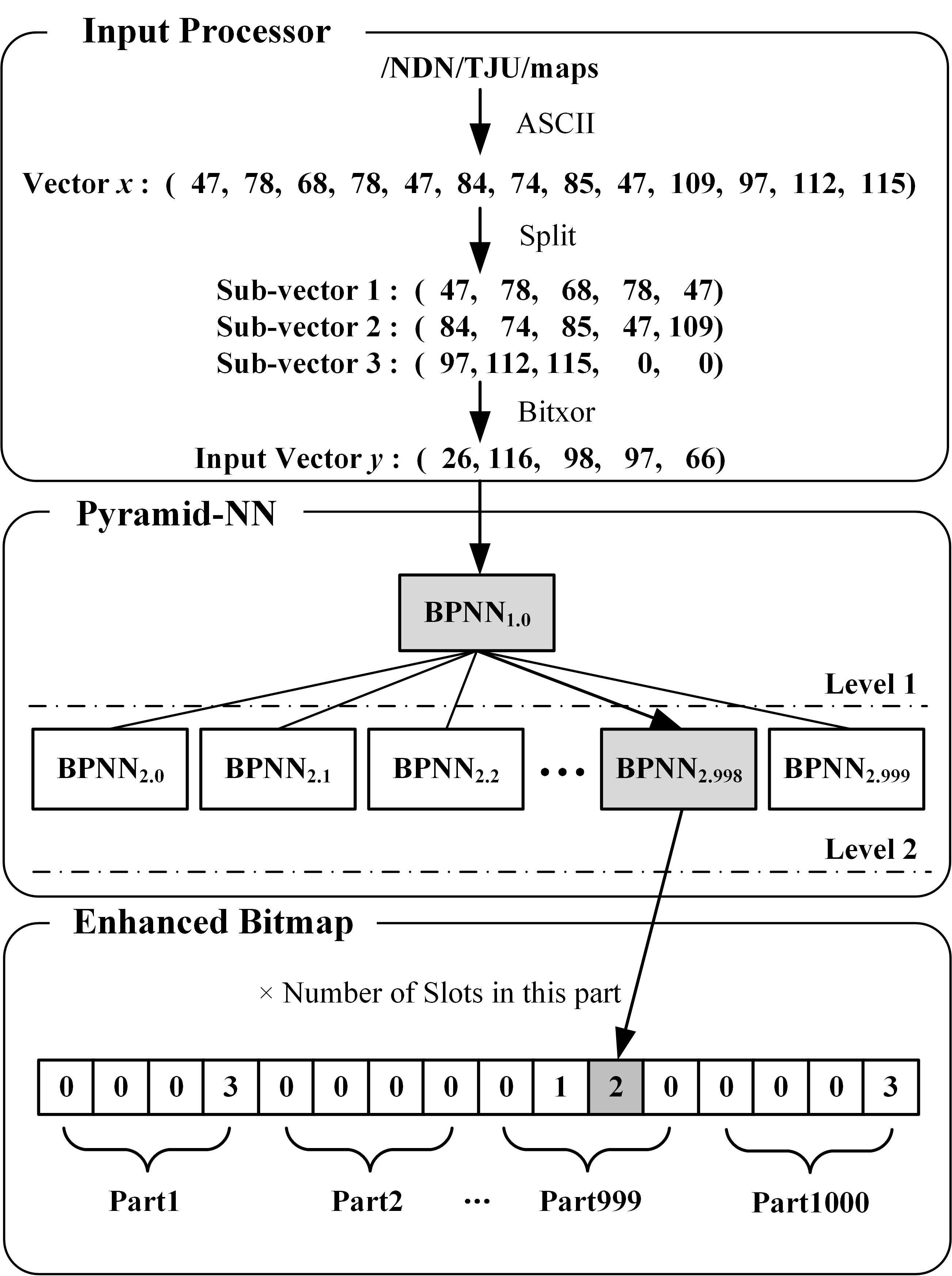}
         \caption{Index architecture of LNI.}
         \label{Index architecture of LNI}
     \end{figure}
 
     \begin{algorithm}[!t]
         \caption{Lookup process of LNI}
         \begin{algorithmic}[1]
         \STATE \textbf{Procedure} Lookup (Name \textit{x})
         \STATE \textit{Segments} $\leftarrow$ \textit{Split}(\textit{x}, 5);
         \STATE \textit{Input} $\leftarrow$ \textit{Bitxor}(\textit{Segments});
         \STATE \textit{L1\_Output} $\leftarrow$ \textit{\rm{BPNN}$_{1.0}$}(\textit{Input});
         \STATE \textit{L2\_Output} $\leftarrow$ \textit{\rm{BPNN}$_{2.L1\_Output}$}(\textit{Input});
         \STATE \textit{Slot} $\leftarrow$ $\lfloor$\textit{L2\_Output} $\times$ Bitmap$_{L1\_Output}$.\textit{size}$\rfloor$;
         \IF{Bitmap[\textit{Slot}] $\ne$ 0}
             \STATE \textit{Part} $\leftarrow$ \textit{Slot} / Part\_Size;
             \STATE \textit{Offset\_Addr} $\leftarrow$ Bitmap[\textit{Slot}];
             \RETURN Base\_Addr[\textit{Part}] + \textit{Offset\_Addr};
         \ENDIF
         \STATE \textbf{End Procedure}
         \end{algorithmic}
         \end{algorithm}
 
 \subsection{Lookup Process}
     The lookup process of LNI is described in Algorithm 2. When a NDN name \textit{x} is input, it is first split, and performed by \textit{Bitxor} operation to get the 5-dimensional input vector (line 2-3). Then it is input into the \textit{\rm{BPNN}$_{1.0}$} for calculation (line 4). Based on \textit{L1\_Output} (i.e., the output of \textit{\rm{BPNN}$_{1.0}$}), \textit{\rm{BPNN}$_{2.L1\_Output}$} is picked and calculated (line 5), and the mapped slot is equal to \textit{L2\_Output} (i.e., the output of \textit{\rm{BPNN}$_{2.L1\_Output}$}) multiplied by the number of slots in corresponding part of the Enhanced Bitmap (line 6). Finally, the mapped slot is queried. If it is not empty, the actual memory address is obtained by adding the offset address recorded in the slot to the base address of memory space corresponding to its part (line 7-11).
 
     An example of name lookup through LNI is indicated by arrow lines in Fig. 7. For an input NDN name \textit{/NDN/TJU/maps}, the ASCII values of every characters form the vector \textit{x}. Then \textit{x} is split every 5 values into sub-vectors (i.e., sub-vector 1 to sub-vector 3), and the corresponding elements in all sub-vectors do the bitxor to obtain the the 5-dimensional input vector (26, 116, 98, 97, 66). Afterwards it is input into Pyramid-NN. In Pyramid-NN, suppose the region number calculated by \textit{\rm{BPNN}$_{1.0}$} is 998, so \textit{\rm{BPNN}$_{2.998}$} is picked next. Calculated by \textit{\rm{BPNN}$_{2.998}$}, the CDF value is got, supposed to be 0.5. Therefore, the predicted slot in the Enhanced Bitmap is equal to 0.5 multiplied by the number of slots in the 999\textit{th} part, namely 0.5 $\times$ 4 = 2. Finally, the slot 3994 in the Enhanced Bitmap is queried. The actual memory address is equal to the base address of memory space corresponding to the 999\textit{th} part plus the offset address 2.
 
     \begin{table}[!t]
    \renewcommand{\arraystretch}{1.3}
    \newcommand{\tabincell}[2]{\begin{tabular}{@{}#1@{}}#2\end{tabular}}
    \centering
    \caption{Training Parameters}
    \label{Training Parameters}
    \setlength{\tabcolsep}{2mm}{
    \begin{tabular}{ccc}
    \hline
    Level                 & level 1           & level 2           \\ \hline
    \# of Epoches         & 20                & 5,000             \\
    Learning Rate         & 0.01              & 0.01              \\
    Loss Function         & Mean Square Error & Mean Square Error \\
    Target Error          & 10$^{-12}$        & 10$^{-12}$        \\
    Minimun Gradient      & 10$^{-12}$        & 10$^{-12}$        \\ \hline
    \end{tabular}}
\end{table}
 
     \begin{figure}[!t]
         \centering
         \includegraphics[width=3.3in]{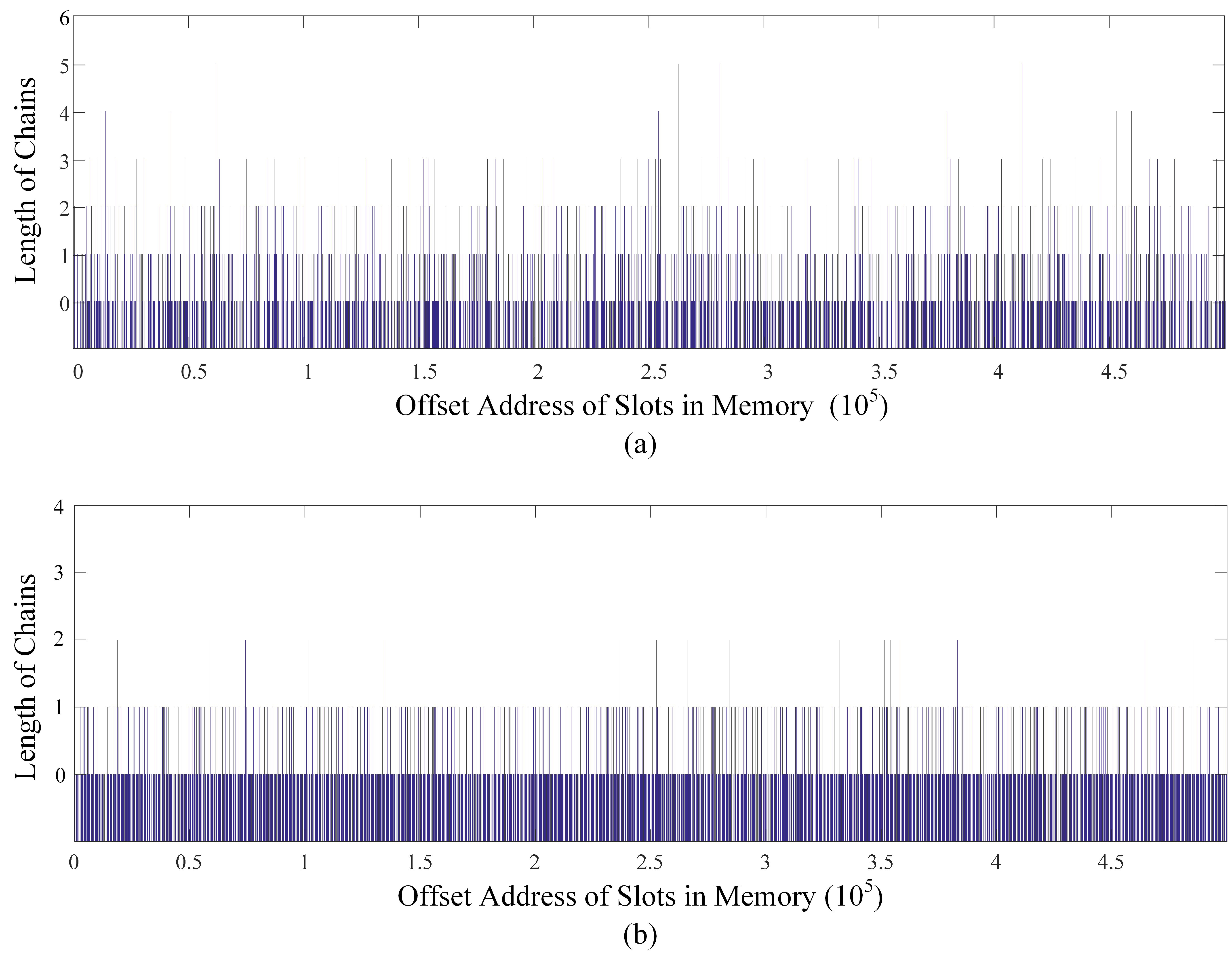}
         \caption{Distribution of mapped slots in memory. (a) CityHash. (b) Pyramid-NN.}
         \label{Distribution of mapped slots in memory}
     \end{figure}
     
     \section{Performance of Pyramid-NN}   
     In this section, Pyramid-NN is compared with some popular hash functions such as MD5 \cite{kirsch2010hash}, CityHash256 \cite{cityhash} and xxHash \cite{xxhash}, as the trained Pyramid-NN is analogous to a hash function. The performance analysis is carried out in four aspects including the memory utilization, the probability of false positive, the model size and the execution speed, which concern if Pyramid-NN is acceptable in practice.
 
     The experimental setup and the dataset are the same as described in Subsection IV(C). Pyramid-NN is implemented to be a two-level model consisting of 1 BPNN at level 1 and 1,000 BPNNs at level 2, where each BPNN has 5 input neurons, 20 hidden neurons and 1 output neuron. More detail parameters are listed in Table III. After training, all the weights and biases are extracted from MATLAB and then Pyramid-NN is regenerated in C++ based on the model specification. The hash functions are also implemented in C++ for fair comparison.
 
     \begin{figure}[!t]
         \centering
         \includegraphics[width=2.4in]{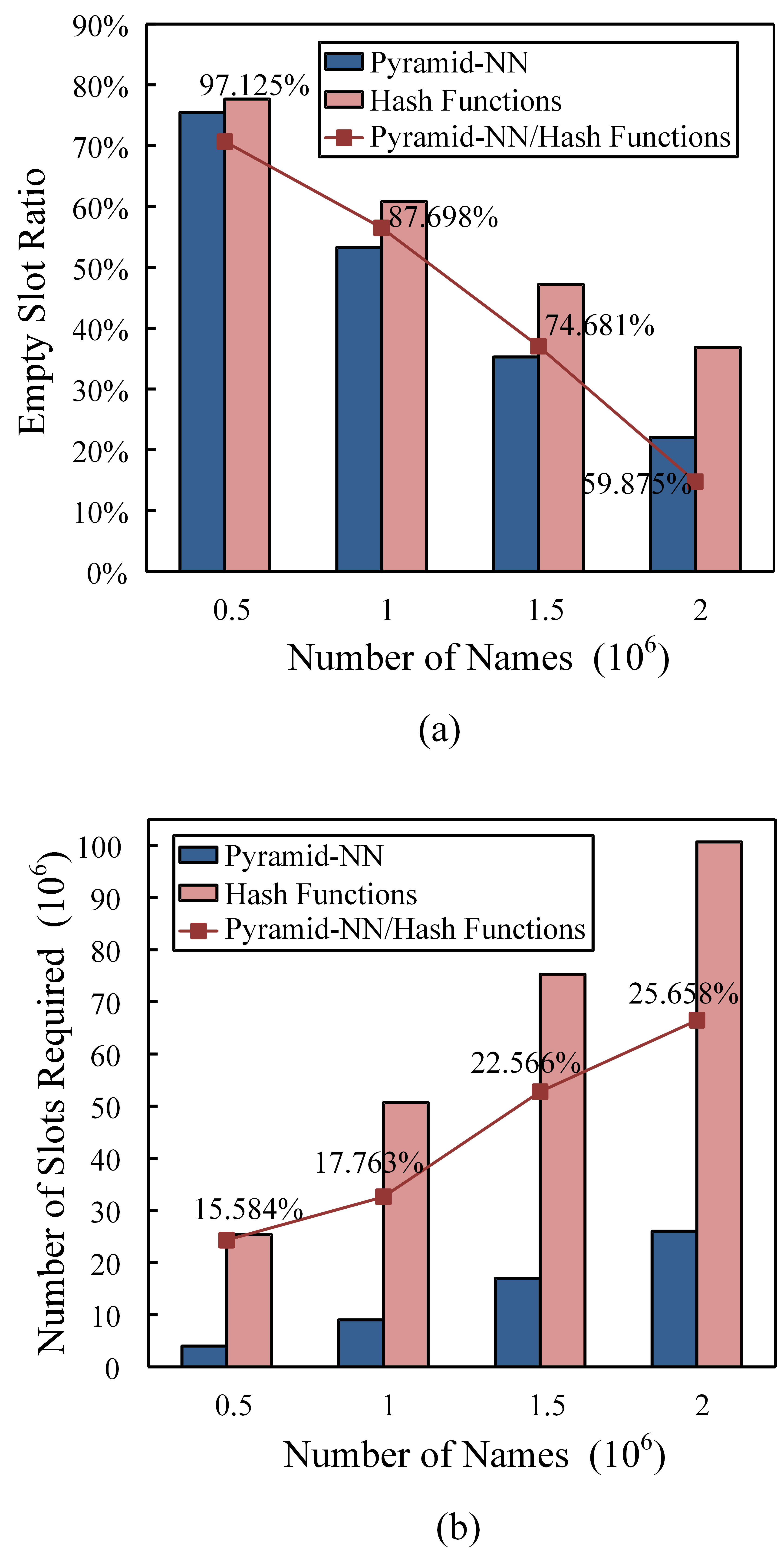}
         \caption{Memory utilization: Pyramid-NN vs. the average of MD5, CityHash256 and xxHash. (a) Empty slot ratio. (b) Number of slots required.}
         \label{Memory utilization performance of Pyramid-NN}
     \end{figure}
 
     \begin{figure}[!t]
         \centering
         \includegraphics[width=2.5in]{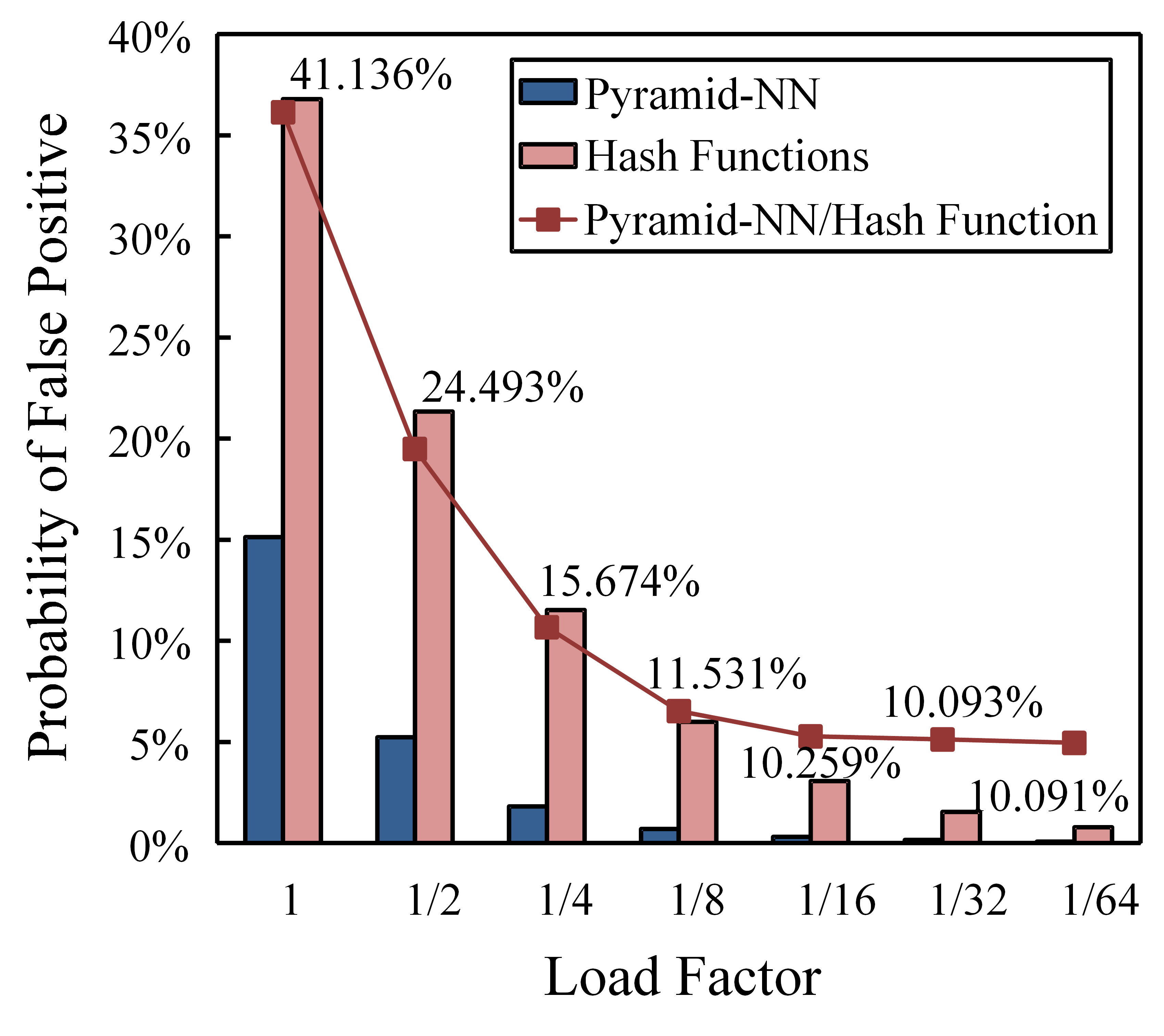}
         \caption{False positive probability performance: Pyramid-NN vs. the average of MD5, CityHash256 and xxHash.}
         \label{False positive probability performance of Pyramid-NN}
     \end{figure}
 
 \subsection{Memory Utilization}
     The distribution of mapped slots in memory is first tested when the load factor is 1 (i.e., 0.5 million names are mapped to 0.5 million slots). As shown in Fig. 8(a), the distribution of mapped slots in memory for CityHash is nonuniform with a large number of conflicts. Hence lots of long chains (the worst case is 5 chains) are required to deal with the conflicts, which significantly impact the lookup speed and memory consumption of the hash table. Instead, as shown in Fig. 8(b), Pyramid-NN maps more uniformly, and the chains required are less and shorter, which means better memory utilization and lookup speed.
 
     \begin{table}[!t]
    \renewcommand{\arraystretch}{1.3}
    \newcommand{\tabincell}[2]{\begin{tabular}{@{}#1@{}}#2\end{tabular}}
    \centering
    \caption{Empty Slot Ratio: Pyramid-NN vs. Hash Functions}
    \label{Table3_Empty Slot Ratio: Pyramid-NN vs. Hash Functions}
    \setlength{\tabcolsep}{2.2mm}{
    \begin{tabular}{ccccc}
    \hline
    \# of Names & Pyramid-NN      & MD5             & CityHash256      & xxHash  \\ \hline
    500,000     & 75.452\%        & 77.020\%        & 78.151\%         & 77.883\%\\
    1,000,000   & 53.301\%        & 60.712\%        & 61.008\%         & 60.614\%\\
    1,500,000   & 35.251\%        & 47.355\%        & 47.011\%         & 47.240\%\\
    2,000,000   & 22.068\%        & 36.692\%        & 37.102\%         & 36.778\%\\ \hline
    \end{tabular}}
\end{table}

     Afterwards, the empty slot ratio (i.e., the proportion of empty slots in the total number of slots in memory) of Pyramid-NN and the hash functions is tested respectively when mapping 0.5-, 1-, 1.5- and 2-million names to 2 million slots. As shown in Fig. 9(a) and Table IV, Pyramid-NN has lower empty slot ratio compared with the hash functions and the gap between the two increases further as the number of names increases. The reason is that the mapping through Pyramid-NN is relatively uniform, while through the hash functions are nonuniform and things get worse as the load factor increases.
 
     Faced with the current network requirement that the packet loss rate should be under 1\% \cite{you2012dipit}, the number of slots required to control the false positive probability under 1\% is tested with 0.5-, 1-, 1.5- and 2-million names. As illustrated in Fig. 9(b) and Table V, the hash functions require 50$\times$ more slots than the input names, but Pyramid-NN only needs about 10$\times$, which makes memory much more efficient. As the number of names increases, the performance of Pyramid-NN declines slightly, for more outputs require Pyramid-NN to have better prediction performance which can be done by increasing the number of BPNNs or levels. However, for the 2 million names, the two-level Pyramid-NN still performs much better than hash functions, where it requires only 25\% of the slots compared to the hash functions.
 
 \subsection{Probability of False Positive}
     Fig. 10 shows the probability of false positive under different load factors with 0.5 million names. Compared with the hash functions, the false positive probability of Pyramid-NN is much lower and the gap between the two increases further as the load factor decreases. More detailed results are given in Table VI. When the load factor is 1/8, the probability of false positive has been reduced to less than 1\% for Pyramid-NN, but still up to about 6\% for the hash functions. The cause lies in that Pyramid-NN achieves more uniform mapping compared with the hash functions (also as illustrated in Fig. 8) which means fewer conflicts and lower false positive probability.
 
     \begin{table}[!t]
    \renewcommand{\arraystretch}{1.3}
    \newcommand{\tabincell}[2]{\begin{tabular}{@{}#1@{}}#2\end{tabular}}
    \centering
    \caption{Number of Slots Required: Pyramid-NN vs. Hash Functions}
    \label{Table4_Number of Slots Required: Pyramid-NN vs. Hash Functions}
    \setlength{\tabcolsep}{1.3mm}{
    \begin{tabular}{ccccc}
    \hline
    \# of Names & Pyramid-NN      & MD5          & CityHash256       & xxHash     \\ \hline
    500,000     & 4,000,000       & 25,000,000   & 26,000,000        & 26,000,000 \\
    1,000,000   & 9,000,000       & 50,000,000   & 51,000,000        & 51,000,000 \\
    1,500,000   & 17,000,000      & 76,000,000   & 75,000,000        & 75,000,000 \\
    2,000,000   & 26,000,000      & 100,000,000  & 102,000,000       & 102,000,000\\ \hline
    \end{tabular}}
\end{table} 
     
 \subsection{Model Size}
     The model size is the size of all the model parameters that need to be stored. As each BPNN consists of 5 input neurons, 20 hidden neurons and 1 output neuron, the weight and bias matrix for input-to-hidden connections in one BPNN is 20 $\times$ 5 and 20 $\times$ 1 in size respectively, while for hidden-to-output connections is 1 $\times$ 20 and 1 $\times$ 1. As all the model parameters are double-precision floating points of 8 bytes, the size of one BPNN is (20 $\times$ 5 + 20 $\times$ 1 + 1 $\times$ 20 + 1 $\times$ 1) $\times$ 8 B = 1,128 B. Further, Pyramid-NN contains 1,001 BPNNs in total, so the total size of Pyramid-NN is 1,128 B $\times$ 1,001 $\approx$ 1.129 MB.
 
     \begin{table}[!t]
    \renewcommand{\arraystretch}{1.3}
    \newcommand{\tabincell}[2]{\begin{tabular}{@{}#1@{}}#2\end{tabular}}
    \centering
    \caption{False Positive Probability: Pyramid-NN vs. Hash Functions}
    \label{Table2_False Positive Probability: Pyramid-NN vs. Hash Functions}
    \setlength{\tabcolsep}{1.3mm}{
    \begin{tabular}{ccccc}
    \hline
    Load Factor & Pyramid-NN    & MD5       & CityHash256  & xxHash   \\ \hline
    1           & 15.131\%      & 36.783\%  & 36.752\%     & 36.812\% \\
    1/2         & 5.224\%       & 21.309\%  & 21.307\%     & 21.371\% \\
    1/4         & 1.806\%       & 11.547\%  & 11.490\%     & 11.537\% \\
    1/8         & 0.691\%       & 5.999\%   & 5.986\%      & 6.003\%  \\
    1/16        & 0.314\%       & 3.063\%   & 3.053\%      & 3.072\%  \\
    1/32        & 0.157\%       & 1.560\%   & 1.547\%      & 1.548\%  \\
    1/64        & 0.078\%       & 0.774\%   & 0.784\%      & 0.749\%  \\ \hline
    \end{tabular}}
    \end{table}

     \begin{table}[!t]
    \renewcommand{\arraystretch}{1.3}
    \newcommand{\tabincell}[2]{\begin{tabular}{@{}#1@{}}#2\end{tabular}}
    \centering
    \caption{Number of Clock Cycles: Pyramid-NN vs. Hash Functions}
    \label{Table5_Number of Clock Cycles: Pyramid-NN vs. Hash Functions}
    \setlength{\tabcolsep}{1mm}{
    \begin{tabular}{ccccc}
    \hline
    \# of Names & Pyramid-NN                & MD5                         & CityHash256              & xxHash      \\ \hline
    500,000     & 2.733 $\times$ 10$^{9}$   & 2.683 $\times$ 10$^{9}$     & 1.060 $\times$ 10$^{9}$  & 2.269 $\times$ 10$^{8}$\\
    1,000,000   & 5.176 $\times$ 10$^{9}$   & 4.935 $\times$ 10$^{9}$     & 1.952 $\times$ 10$^{9}$  & 4.760 $\times$ 10$^{8}$\\
    1,500,000   & 7.644 $\times$ 10$^{9}$   & 7.167 $\times$ 10$^{9}$     & 2.802 $\times$ 10$^{9}$  & 7.031 $\times$ 10$^{8}$\\
    2,000,000   & 10.188 $\times$ 10$^{9}$  & 9.592 $\times$ 10$^{9}$     & 3.665 $\times$ 10$^{9}$  & 9.173 $\times$ 10$^{8}$\\ \hline
    \end{tabular}}
\end{table} 
 
     \begin{table*}[!t]
\renewcommand{\arraystretch}{1.3}
\newcommand{\tabincell}[2]{\begin{tabular}{@{}#1@{}}#2\end{tabular}}
\centering
\caption{Memory Consumption (MB): LNI-FIB vs. Other Indexes}
\label{Table6_Memory Consumption (MB): LNI vs. Other Indexes}
\setlength{\tabcolsep}{2mm}{
\begin{tabular}{ccccccc}
\hline
\# of Names & LNI-FIB    & Binary Patricia Trie-FIB & HT with MD5-FIB  & HT with CityHash256-FIB & HT with xxHash-FIB & B-MaFIB    \\ \hline
500,000 & \multirow{4}{30pt}{\centering58.258} & 16.265 & \multirow{4}{30pt}{\centering400.000} &\multirow{4}{30pt}{\centering408.000} & \multirow{4}{30pt}{\centering408.000} & \multirow{4}{30pt}{\centering2,308.097}  \\
1,000,000 & & 32.507 & & & & \\
1,500,000 & & 48.762 & & & & \\
2,000,000 & & 65.046 & & & & \\ \hline
\end{tabular}}
\end{table*}
 
 \subsection{Execution Speed}
     The execution speed is tested against 0.5-, 1-, 1.5- and 2-million names for Pyramid-NN and the hash functions. As listed in Table VII, thanks to the small enough size of Pyramid-NN, the number of clock cycles taken with 0.5-, 1-, 1.5- and 2-million names is only 2.733 $\times$ 10$^{9}$, 5.176 $\times$ 10$^{9}$, 7.644 $\times$ 10$^{9}$ and 10.188 $\times$ 10$^{9}$ respectively, which is an acceptable high speed on the same order of magnitude as that of the hash functions. Therefore, it is feasible to use Pyramid-NN in the index design for NDN forwarding plane, which can satisfy the current network requirements for fast packet processing.
 
     \section{Evaluation and Discussion}
     In the section, the performance of LNI-based FIB, called LNI-FIB, is evaluated in terms of the memory consumption, the probability of false positive and the throughput. These results are compared with HT-FIB \cite{kirsch2010hash}, the admitted efficient Binary Patricia Trie-FIB \cite{song2015scalable} and Bloom Filter-based B-MaFIB \cite{Li20185G}.
 
     \begin{table}[!t]
\renewcommand{\arraystretch}{1.3}
\newcommand{\tabincell}[2]{\begin{tabular}{@{}#1@{}}#2\end{tabular}}
\centering
\caption{Memory Devices in a Line Card \cite{song2015scalable}}
\label{Table7_Memory Devices in a Line Card}
\setlength{\tabcolsep}{2mm}{
\begin{tabular}{ccccc}
\hline
Device & Single-chip Size & \#  & Total Size       & Lookup Time \\ \hline
TCAM   & 2.384 MB         & 4   & 9.537 MB         & 2.7 ns      \\
SRAM   & 32.187 MB        & 4   & 128.746 MB       & 0.47 ns     \\
DRAM   & 3.725 GB         & 4-8 & 14.901-29.802 GB & 50 ns       \\ \hline
\end{tabular}}
\end{table} 
 
     \begin{table*}[!t]
    \renewcommand{\arraystretch}{1.3}
    \newcommand{\tabincell}[2]{\begin{tabular}{@{}#1@{}}#2\end{tabular}}
    \centering
    \caption{Probability of False Positive: LNI-FIB vs. Other Indexes}
    \label{Table8_Probability of False Positive: LNI vs. Other Indexes}
    \setlength{\tabcolsep}{4mm}{
    \begin{tabular}{cccccc}
    \hline
    \# of Names & LNI-FIB    & HT with MD5-FIB  & HT with CityHash256-FIB & HT with xxHash-FIB & B-MaFIB   \\ \hline
    500,000     & 0.078\%    & 0.774\%          & 0.784\%                 & 0.750\%            & 2.371\% \\
    1,000,000   & 0.243\%    & 1.553\%          & 1.557\%                 & 1.560\%            & 4.598\% \\
    1,500,000   & 0.488\%    & 2.316\%          & 2.319\%                 & 2.310\%            & 6.795\% \\
    2,000,000   & 0.817\%    & 3.061\%          & 3.069\%                 & 3.080\%            & 8.925\% \\ \hline
    \end{tabular}}
    \end{table*} 
 
     \begin{table*}[!t]
    \renewcommand{\arraystretch}{1.3}
    \newcommand{\tabincell}[2]{\begin{tabular}{@{}#1@{}}#2\end{tabular}}
    \centering
    \caption{Throughput (MSPS): LNI-FIB vs. Other Indexes}
    \label{Table9_Throughput (MSPS): LNI vs. Other Indexes}
    \setlength{\tabcolsep}{2mm}{
    \begin{tabular}{ccccccc}
    \hline
    \# of Names & \begin{tabular}[c]{@{}c@{}}LNI-FIB\\ (SRAM)\end{tabular} & \begin{tabular}[c]{@{}c@{}}Binary Patricia Trie-FIB\\ (SRAM)\end{tabular} & \begin{tabular}[c]{@{}c@{}}HT with MD5-FIB\\ (DRAM)\end{tabular} & \begin{tabular}[c]{@{}c@{}}HT with CityHash256-FIB\\ (DRAM)\end{tabular}& \begin{tabular}[c]{@{}c@{}}HT with xxHash-FIB\\ (DRAM)\end{tabular} & \begin{tabular}[c]{@{}c@{}}B-MaFIB\\ (SRAM + DRAM)\end{tabular} \\ \hline
    500,000     & 162.16                                                      & 134.75                                                                & 1.55                                                                                                                 & 3.78       & 14.26                                                       & 6.88                                                          \\
    1,000,000   & 173.37                                                      & 142.84                                                                & 1.71                                                                                                                  & 4.23        & 15.57                                                      & 7.53                                                          \\
    1,500,000   & 176.90                                                      & 143.24                                                                & 1.77                                                                                                               & 4.46         & 16.52                                                     & 8.51                                                          \\
    2,000,000   & 177.37                                                      & 139.87                                                                & 1.77                                                                                                            & 4.58         & 17.26                                                     & 8.72                                                          \\ \hline
    \end{tabular}}
    \end{table*} 
 
 \subsection{Experimental Setup}
   The experimental setup and the dataset are the same as described in Subsection IV(C). All the index schemes are implemented in C++ for a fair comparison.
 
     For LNI-FIB, it is composed of two LNIs \cite{Li20185G}. The hyperparameter setting and generation of Pyramid-NN is the same as described in Section IV, while the size of each slot in the Enhanced Bitmap is set to 2 bytes. For Binary Patricia Trie-FIB, one entry has 4 bytes which is a pointer to access the memory storing actual packet information. For HT-FIB, it is implemented with MD5, CityHash256 and xxHash as the hash function respectively, while one entry also has 4 bytes. For B-MaFIB, the size of Bloom filter is 2$^{24}$ bits and the size of MA is 24 bits, while the size of each slot in the Bitmap is 2 bytes.
 
 \subsection{Memory Consumption}
     If the index can be stored in small and fast memories (e.g., SRAM), the routers will easily complete fast packet forwarding. To determine which indexes can be deployed on SRAMs, the memory consumption of LNI-FIB, Binary Patricia Trie-FIB, HT-FIB and B-MaFIB is compared and analyzed. Given that LNI-FIB, HT-FIB and B-MaFIB are static, the memory consumption of them is evaluated under the condition of 1\% false positive probability for all testing sets. For Binary Patricia Trie-FIB, the memory consumption of it is evaluated with different number of names.
 
   As shown in Fig. 11, the memory consumption of LNI-FIB is 58.258 MB, which is 10\% less than that of Binary Patricia Trie-FIB when the number of names is 2 million. And as for HT-FIB and B-MaFIB, they map data to slots in memory randomly and have a large number of conflicts which result in more memory consumption to reduce the conflicts. In contrast, compared with HT-FIB and B-MaFIB, the LNI-FIB with higher memory utilization significantly decreases the memory consumption by 85\% and 97\%, respectively.
 
   More detailed results are given in Table VIII. The memory consumption of LNI-FIB includes the model parameters of Pyramid-NN and the slots in the Enhanced Bitmap. First, the model parameters of Pyramid-NN consume 1.129 MB as indicated in Subsection VI(C). Thus the memory consumption of two Pyramid-NNs is 1.129 MB $\times$ 2 = 2.258 MB. Further, under the condition of 1\% false positive probability, the total number of slots required in two Enhanced Bitmaps is 28-million, which consume 56 MB. Consequently, the on-chip memory consumption of LNI-FIB is 58.258 MB. As listed in Table IX, a line card can be configured with four channels of 32.187 MB single-chip SRAMs for a total of 128.746 MB in size \cite{song2015scalable}. Table VIII shows the memory consumption in megabytes, LNI-FIB can easily fit into contemporary SRAMs in commercial line card, as the memory consumption of it is less than 128.746 MB. In contrast, the memory consumption of HT-FIB and B-MaFIB limits its deployment on SRAMs.
 
     \begin{figure}[!t]
         \centering
         \includegraphics[width=3.3in]{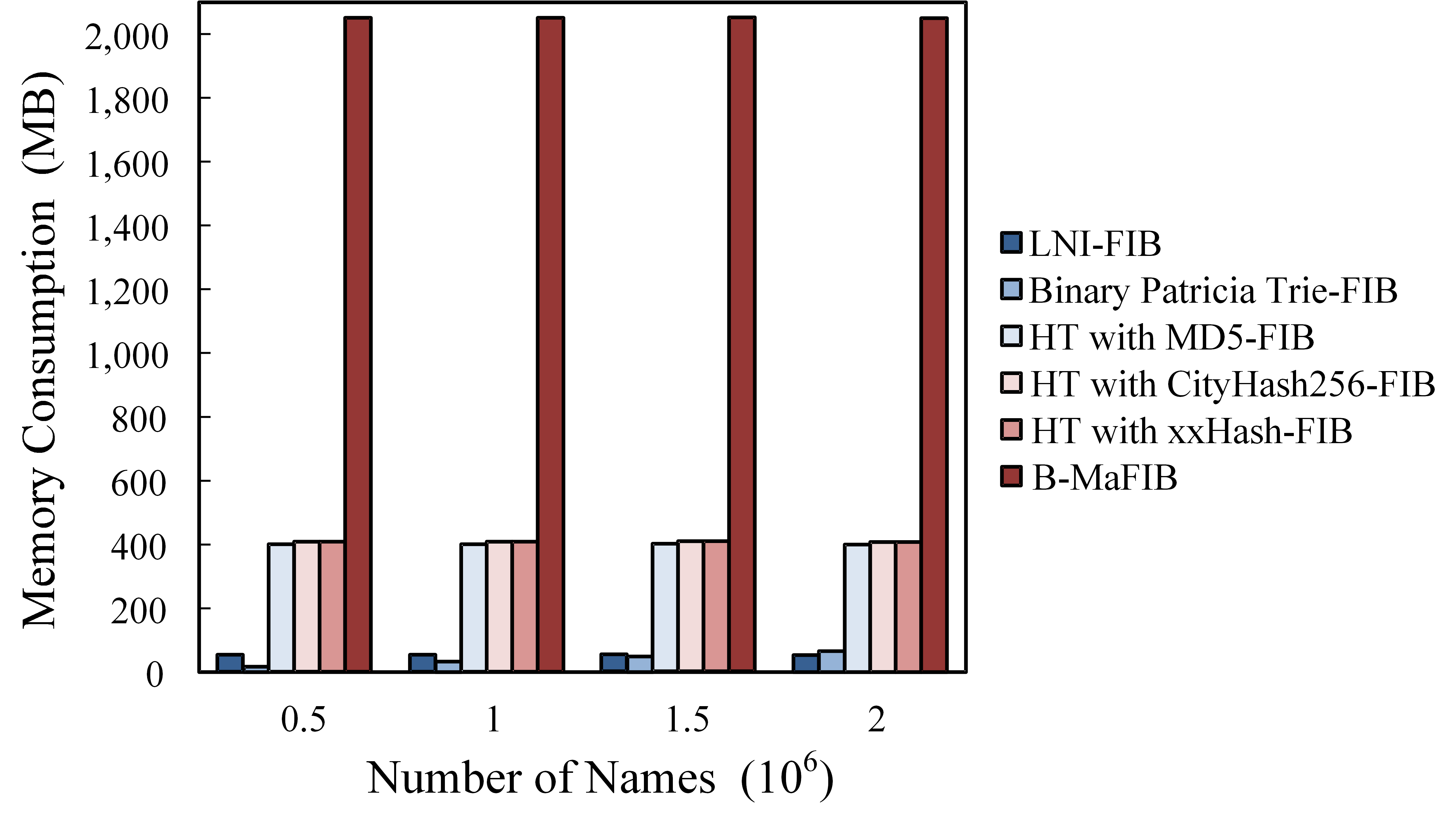}
         \caption{Memory consumption (MB): LNI-FIB vs. other indexes.}
         \label{Memory consumption (MB): LNI-FIB vs. other indexes}
     \end{figure}
 
 \subsection{Probability of False Positive}
     Given the current network requirements that the packet loss rate should be under 1\% \cite{you2012dipit}, the false positive probability of LNI-FIB, HT-FIB and B-MaFIB with 32 million slots is compared and analyzed in different number of names.
 
   As shown in Fig. 12, the false positive probability of LNI-FIB is 0.817\% as the number of names is 2 million, which is less than a third and a ninth of that of HT-FIB and B-MaFIB, respectively. The reason is that LNI-FIB maps data to slots more uniformly, but as for the other, the mapping is relatively nonuniform and has a large number of conflicts.
 
     More detailed results are given in Table X. The false positive probability of LNI-FIB is approximately equal to 0.078\%, 0.243\%, 0.488\% and 0.817\% as the number of names is 0.5-, 1-, 1.5- and 2-million, which is lower than 1\%, meeting the current network requirements. In comparison, the false positive probability of HT-FIB and B-MaFIB is much higher than that of LNI-FIB.
 
 \subsection{Throughput}
     Faced with the current network requirement for fast packet processing, the throughput in execution of LNPM is tested against 0.5-, 1-, 1.5- and 2-million names for LNI-FIB, Binary Patricia Trie-FIB, HT-FIB and B-MaFIB.
 
   The comparison is shown in Fig. 13, LNI-FIB outperforms the others in throughput due to the deployment on SRAMs. For 2 million names, the throughput of LNI-FIB is 177.37 MSPS, which is about 26\% more than that of Binary Patricia Trie-FIB. The reason is that each traversing from one level to the next one in Binary Patricia Trie-FIB requires one memory access and that the average height of it is much higher than traditional trie reduces its lookup speed. And compared with HT-FIB and B-MaFIB, the throughput of LNI-FIB is about 100$\times$, 38$\times$, 10$\times$ and 20$\times$ more than that of HT with MD5-FIB, HT with CityHash256-FIB,HT with xxHash-FIB and B-MaFIB respectively. Because HT-FIB and B-MaFIB have large footprint, so that they have to be deployed on DRAM, which significantly reduces its throughput. Thus, HT-FIB and B-MaFIB cannot well meet the current network requirement for packet processing.
 
     More detailed experimental results are given in Table XI. For 0.5-, 1-, 1.5- and 2-million names, the throughput of LNI-FIB provides a higher throughput of about 162.16 MSPS, 173.37 MSPS, 176.90 MSPS and 177.37 MSPS, as multi-level Pyramid-NN in LNI-FIB consists of simple BPNNs with small size and can run in parallel, which can be executed fast in NDN forwarding plane.
 
     \begin{figure}[!t]
         \centering
         \includegraphics[width=2.55in]{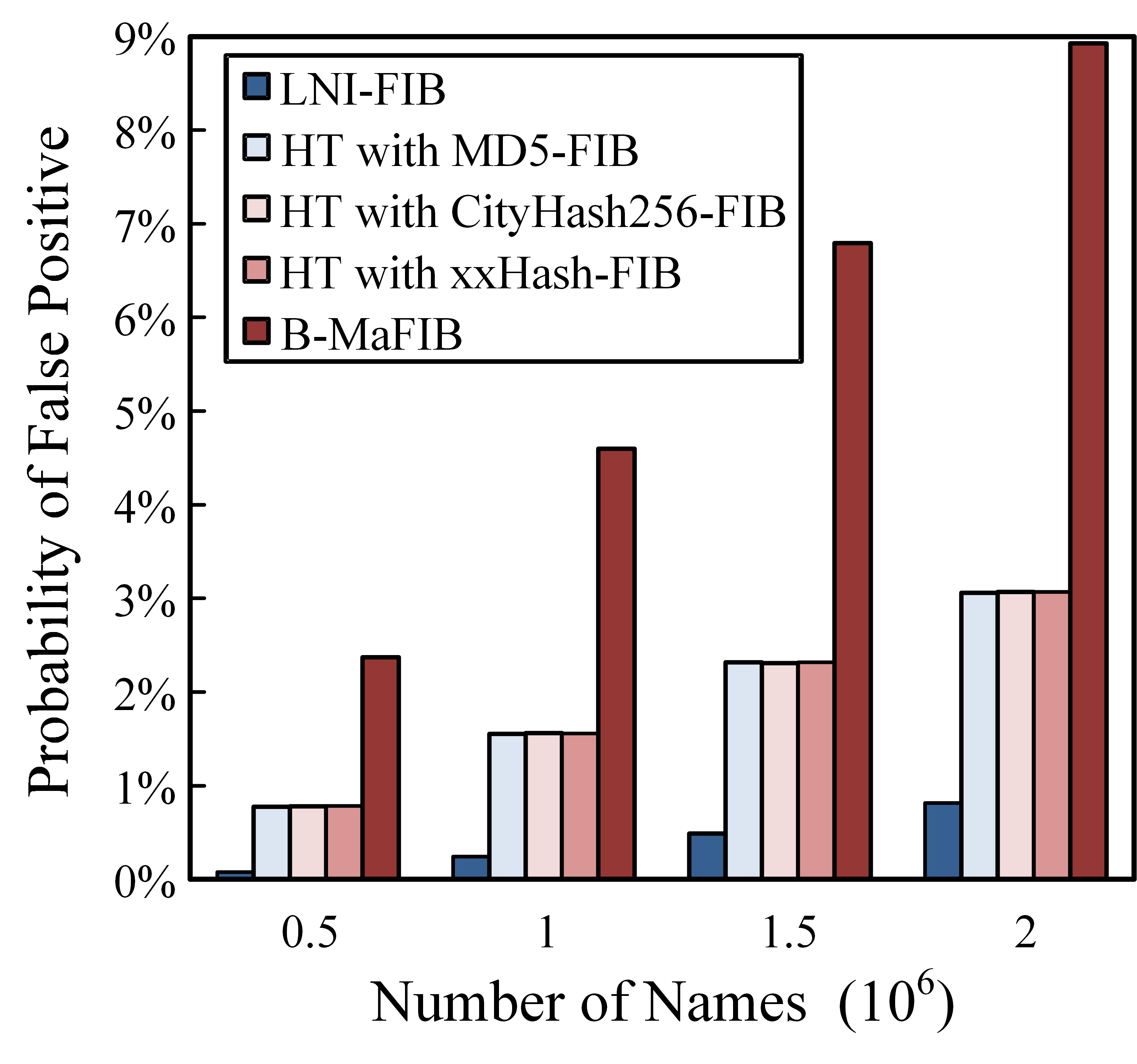}
         \caption{Probability of false positive: LNI-FIB vs. other indexes.}
         \label{Probability of false positive: LNI-FIB vs. other indexes}
     \end{figure}
 
    \begin{figure}[!t]
         \centering
         \includegraphics[width=3.3in]{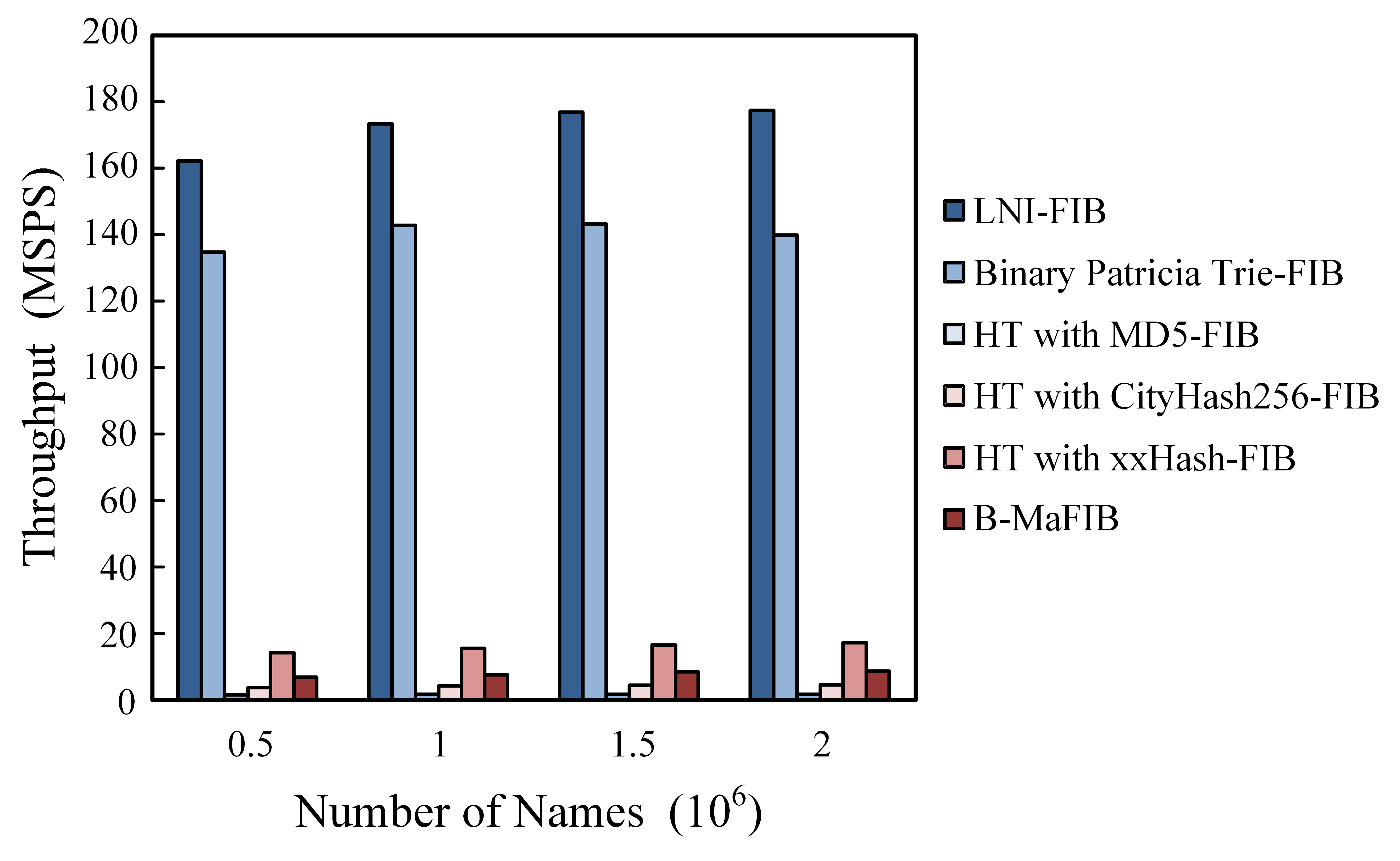}
         \caption{Throughput (MSPS): LNI-FIB vs. other indexes.}
         \label{Throughput (MSPS): LNI-FIB vs. other indexes}
     \end{figure}
 
 \subsection{Discussion}
     As LNI can learn the distributions of name retrieved in the static memory, LNI-FIB can reduce the memory consumption and the probability of false positive to 58.258 MB and 0.817\% respectively for 2 million names. And because it can be deployed on SRAMs, the throughput is about 177 MSPS, which is much better than that of HT-FIB and B-MaFIB. More importantly, LNI-FIB can adapt to the distributions of real NDN names in the static memory by retraining with the names, which will not only reduce the memory consumption and the probability of false positive, but also ensure the performance of real name lookup.
 
     \section{Conclusion and Future work}
     This paper proposed a smart mapping model via neural networks called Pyramid-NN  to learn the distributions of names retrieved in the memory. Based on Pyramid-NN, an index named LNI was proposed. Through learning the distributions of the names retrieved in the static memory, LNI that will be trained by real NDN names offline and preset in content routers in the future can not only reduce the memory consumption and the probability of false positive, but also ensure the performance of real name lookup. The performance of LNI-FIB is evaluated and the experimental results show that its performance in terms of memory consumption, false positive probability and throughput can be significantly improved by utilizing neural network, which can meet the current network requirement well for fast packet processing.  
     
     One promising future direction would be to extend this design to the build of an engine running on multiple parallel threads with real packets. And another is exploring more efficient neural networks, which can not only realize the mapping function, but also map the real NDN names dynamically. 


%





\ifCLASSOPTIONcaptionsoff
  \newpage
\fi



\bibliographystyle{IEEEtran}
\bibliography{IEEEabrv,ref}
%



%

\begin{IEEEbiography}[{\includegraphics[width=1in,height=1.25in,clip,keepaspectratio]{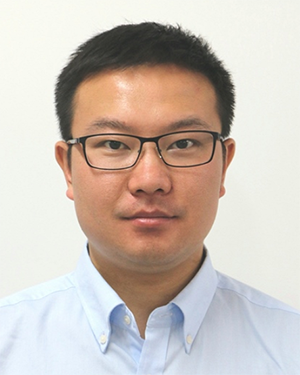}}]{Zhuo Li}
received Ph.D. degree from Tianjin University, China in 2015 and M.S. degree from East China Normal University, China in 2010. Before 2017, he spent one year as a Post-Doctoral Research Associate in the Computer Science Department at the University of Arizona. He is currently working in the School of Microelectronics at Tianjin University. His main research interests include router architecture, fast packet processing, wireless communications and Future Internet Architecture, e.g., Named Data Networking. He now serves as the reviewer for IEEE Intelligent Transportation Systems Transactions, IEEE Communications Magazine, IEEE Access, IEEE Transactions on Mobile Computing and IEEE Communications Letters.
\end{IEEEbiography}

\begin{IEEEbiography}[{\includegraphics[width=1in,height=1.25in,clip,keepaspectratio]{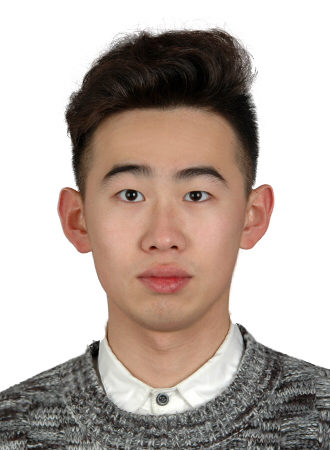}}]{Jindian Liu}
is a master candidate in the School of Microelectronics, Tianjin University. He got the B.S. degree in electronic information science and technology from Yanshan University, China, in 2019. He research interests mainly include router architecture, fast packet processing and Future Internet Architecture, e.g., Named Data Networking.
\end{IEEEbiography}

\begin{IEEEbiography}[{\includegraphics[width=1in,height=1.25in,clip,keepaspectratio]{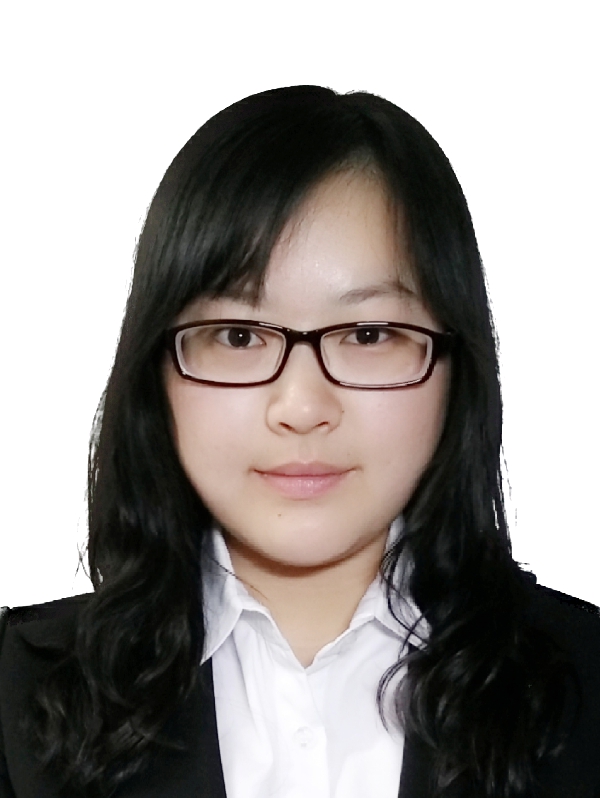}}]{Liu Yan}
is a master candidate in the School of Microelectronics, Tianjin University. She got the B.S. degree in electronic information engineering (English intensive) from Dalian University of Technology, China, in 2017. Her research interests mainly include router architecture, fast packet processing and Future Internet Architecture, e.g., Named Data Networking.
\end{IEEEbiography}

\begin{IEEEbiography}[{\includegraphics[width=1in,height=1.25in,clip,keepaspectratio]{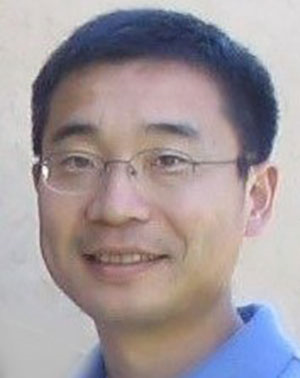}}]{Beichuan Zhang}
received Ph.D. degree from UCLA in 2003, and B.S. degree from Peking University, China in 1995. Before 2005, he spent two years as a postdoc at USC/ISI-East and UCLA. He is currently working in the Computer Science Department at the University of Arizona. His research area is computer networks in general. He has been working on Internet routing architectures and protocols, green networks, routing security, and Internet content distribution. Dr. Zhang is the receipient of the Applied Networking Research Prize by ISOC and IRTF in 2011, and the Best Paper Award at ICDCS in 2005.
\end{IEEEbiography}

\begin{IEEEbiography}[{\includegraphics[width=1in,height=1.25in,clip,keepaspectratio]{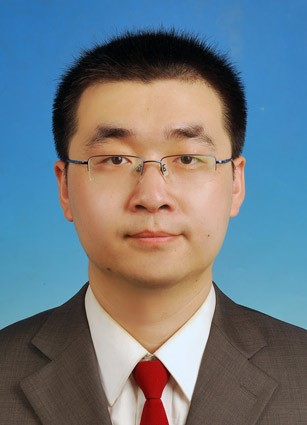}}]{Peng Luo}
  received Ph.D. degree from Tianjin University, China in 2012. He is currently working in the State Grid Hebei Electric Power Research Institute. His main research interests include power system auto-mation, new energy power system and intel-ligent power distribution technology.
\end{IEEEbiography}

\begin{IEEEbiography}[{\includegraphics[width=1in,height=1.25in,clip,keepaspectratio]{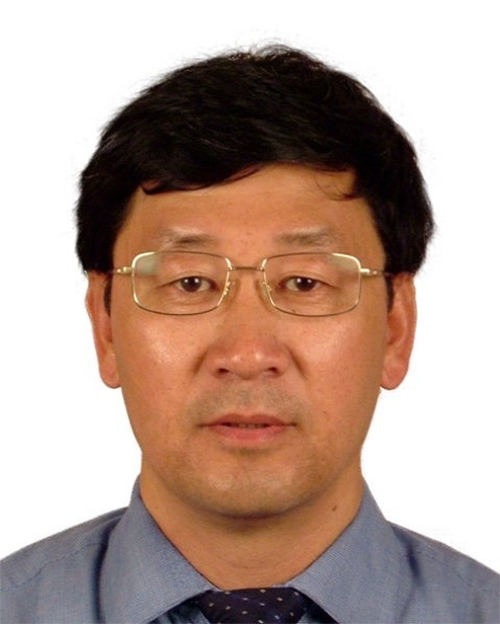}}]{Kaihua Liu}
is currently a full professor with the School of Microelectronics, Tianjin University, China. He received the B.S. degree in Semiconductor physics and devices, the M.S. degree in circuits and systems and Ph.D. degree in signal
and information processing from Tianjin University, in 1981, 1991 and 1999, respectively. His current research interests include radio frequency identification, digital signal processing and Future Internet Architecture. He is a member of IEEE.
\end{IEEEbiography}








\end{document}